\documentclass[aps,prb,twocolumn,superscriptaddress,groupedaddress,showpacs]{revtex4}
\usepackage{graphicx,xcolor}
\usepackage{subfigure}

\begin{document}

\title{Dual-fermion approach to interacting disordered fermion systems}

\author{S.-X. Yang}
\email{yangphysics@gmail.com}
\affiliation{Department of Physics and Astronomy, Louisiana State University, Baton Rouge, Louisiana 70803, USA}
\affiliation{Center for Computation and Technology, Louisiana State University, Baton Rouge, Louisiana 70803, USA}
\author{P.\ Haase}
\affiliation{Department of Physics, University of G\"ottingen, 37077 G\"ottingen, Germany}
\author{H. Terletska}
\affiliation{Department of Physics and Astronomy, Louisiana State University, Baton Rouge, Louisiana 70803, USA}
\affiliation{Condensed Matter Physics and Materials Science Department, Brookhaven National Laboratory, Upton, New York 11973, USA }
\author{Z. Y. Meng}
\affiliation{Department of Physics and Astronomy, Louisiana State University, Baton Rouge, Louisiana 70803, USA}
\affiliation{Center for Computation and Technology, Louisiana State University, Baton Rouge, Louisiana 70803, USA}
\author{T.\ Pruschke}
\affiliation{Department of Physics, University of G\"ottingen, 37077 G\"ottingen, Germany}
\author{J.\ Moreno}
\affiliation{Department of Physics and Astronomy, Louisiana State University, Baton Rouge, Louisiana 70803, USA}
\affiliation{Center for Computation and Technology, Louisiana State University, Baton Rouge, Louisiana 70803, USA}
\author{M.\ Jarrell}
\affiliation{Department of Physics and Astronomy, Louisiana State University, Baton Rouge, Louisiana 70803, USA}
\affiliation{Center for Computation and Technology, Louisiana State University, Baton Rouge, Louisiana 70803, USA}

\date{\today}

\begin{abstract}
We generalize the recently introduced dual fermion (DF) formalism for disordered fermion systems
by including the effect of interactions. For an interacting disordered system the contributions to the
full vertex function have to be separated into crossing-asymmetric and crossing-symmetric scattering processes, and addressed
differently when constructing the DF diagrams. By applying our approach to the Anderson-Falicov-Kimball model 
and systematically restoring the nonlocal correlations
in the DF lattice calculation, we show a significant improvement 
over the Dynamical Mean-Field Theory and the Coherent Potential Approximation
for both one-particle and two-particle quantities.

\end{abstract}


\maketitle

\section{Introduction}
The transport and thermodynamic properties of many real materials are strongly influenced by disorder and strong electron 
correlations~\cite{Lee_RevModPhys,Belitz_RevModPhys}. The interplay of these two effects can lead to many interesting 
novel phenomena.  In particular, both disorder and electron-electron interactions are known to be the driving mechanisms 
for metal-insulator transitions, although of different nature. Electron correlations induce the Mott-Hubbard metal-insulator transition
with the opening of a gap in the single particle excitation spectrum~\cite{Mott}. On the other hand, coherent back-scattering 
of electrons off disorder-induced non-periodic potentials can lead to their localization, known as Anderson 
localization~\cite{Anderson}. Despite intensive studies, the proper modeling of disordered interacting systems remains 
a great challenge.  

Mean field methods like the coherent potential approximation (CPA)~\cite{leath66,p_soven_67,d_taylor_67,shiba71} and the 
dynamical mean field theory (DMFT)~\cite{DMFA1,DMFA2,DMFA3,DMFA_reviews1,DMFA_reviews2}  
have revolutionized the study of disordered and correlated systems.  
These are single-site mean field approximations with an averaged local momentum-independent effective medium. As 
single-site methods, both the CPA and DMFT fail to take into account non-local correlations, which are found to be 
important in many cases. For example, in correlated clean systems one frequently observes ordered states with 
non-local order parameters which cannot be accounted for within the DMFT. Likewise, for non-interacting disordered systems 
it is well known that the CPA, while being rather successful describing electronic structures, completely fails to 
capture Anderson localization~\cite{Lee_RevModPhys}. 

There have been a number of attempts to develop systematic nonlocal extensions to the CPA and DMFT. These include
such cluster extensions as the Molecular Coherent Potential Approximation (MCPA) \cite{tsukada69,tsukada72,f_ducastelle_74}, 
Dynamical Cluster Approximation (DCA) \cite{Hettler98,Hettler00,m_jarrell_01a}, Cluster Coherent Potential 
Approximation (CCPA) \cite{mookerjee73,kaplan76a,kaplan76b,kumar82,mookerjee87}, and the Traveling Cluster 
Approximation (TCA) \cite{mills78,kaplan80}.  These methods generally extend the CPA and DMFT by replacing the single-site 
impurity problem by that of a finite-size cluster coupled to a mean field bath.

A distinctly different approach called the Dual Fermion (DF) method has been developed to incorporate non-local correlations 
introduced by both disorder and interactions.  Originally constructed for interacting clean systems\cite{Rubtsov08},
it has been recently extended to study disordered non-interacting electronic systems~\cite{h_terletska_13} 
and disordered dipole system~\cite{Osipov_13}. 
Note that earlier a very similar idea using parquet method has also been proposed by Janis~\cite{Janis_parquet},
though DF method is more elegant, systematic, and most importantly, it designs an exact mapping from real fermion lattice 
onto dual fermion lattice.
In this work, we extend the dual fermion approach further so both disorder and interaction effects can be be treated on equal footing.
By separating the scattering vertex contributions into crossing-asymmetric and crossing-symmetric components, we manage to derive 
the proper DF mapping and construct the DF Feynman diagrams, which are now more complicated due to the different 
scattering processes arising from disorder and Coulomb interaction, respectively.  We apply the method to the 
Anderson-Falicov-Kimball model.  Our numerical results for 1D systems show a remarkable correction to the 
DMFT-CPA results and are consistent with DCA calculations of large clusters. Finally, the phase diagram for 2D 
systems is determined by using both one- and two-particle quantities.

The paper is organized as follows: in Section II we describe the details of the DF formalism for treating both 
disorder and electron-electron interactions.  Results for one- and two-particle properties obtained from applying 
our DF formalism to the Anderson-Falicov-Kimball Model and how they compare with DMFT-CPA data are presented in
Sec.\ III. Section IV summarizes and concludes the paper.

\section{Formalism}

\subsection{Dual fermion mapping}

The paradigm for studying disordered correlated systems is the Anderson-Hubbard model
\begin{equation}
\mathcal{H} = 
 \sum_{{\bf k},\sigma} (\epsilon_k-\mu) c_{{\bf k}\sigma}^{\dagger}c_{{\bf k}\sigma}
- \sum_{i\sigma}v_{i}n_{i\sigma}
+ U\sum_{i}n_{i\uparrow}n_{i\downarrow}
\label{eq:AH} %
\end{equation}
where $\epsilon_k$ is the dispersion of the band electrons, $\mu$ is the chemical potential, 
$U$ is the Coulomb interaction and the on-site disorder potential $v_i$ is distributed 
according to some given probability density ${\cal P}(v_i)$. The latter can have in principle any form, but
for the present purpose we specify it as
\begin{equation} 
\mathcal{P}(v_{i})=\Theta(D/2-|v_{i}|)/D, 
\label{eq:distribution}
\end{equation}
where $\Theta(x)$ is the step function
\begin{equation} 
\Theta(x) = \left\{
\begin{array}{c l}
  1, & x \ge 0 \\
  0, & x < 0
\end{array}
\right.,
\end{equation}
and $D$ is the disorder strength.

Following the derivation of the dual fermion mapping for the non-interacting disordered case~\cite{h_terletska_13},
after introducing the auxiliary dual fermion degrees of freedom and then integrating out the real fermion degrees 
of freedom (see appendix for details), one arrives at an effective action 
(to simplify the notation we represent the fermionic Matsubara frequency $iw_n$ as $w$ and
bosonic Matsubara frequency $i\nu_m$ as $\nu$ in the following)
\begin{equation}
S[f,f^{*}]=-\sum_{w,{\bf k}}G_{d0}^{-1}(w,{\bf k})f_{w,{\bf k}}^{*}f_{w,{\bf k}}+\sum_i V_{d,i}
\end{equation}
in which the bare dual Green function is defined as the difference of lattice Green function $G_{lat}$ 
and impurity Green function $G_{imp}$
\begin{equation}
G_{d0}(w,{\bf k})\equiv G_{lat}(w,{\bf k})-G_{imp}(w)\end{equation}
and the dual potential (by keeping only the lowest two-body interaction)
\begin{eqnarray}
 V_{d,i} & = & \frac{1}{2}\sum_{w,w^{\prime}}V^{p,0}(w,w^{\prime})
f_{i,w}^{*}f_{i,w^{\prime}}^{*}f_{i,w^{\prime}}f_{i,w}
\nonumber \\
 & + & \frac{1}{4}\sum_{w,w^{\prime},\nu}V^{p,1}(\nu)_{w,w^{\prime}}
f_{i,w+\nu}^{*}f_{i,-w}^{*}f_{i,-w^{\prime}}f_{i,w^{\prime}+\nu}
\end{eqnarray}
is split into crossing-asymmetric, with interaction strength $V^{p,0}$, and crossing-symmetric terms, with interaction strength $V^{p,1}$.
They are parametrized by the real fermion full vertex  
\begin{eqnarray}
 V^{p,0} & = & F^{p,0}
\nonumber \\
  V^{p,1} & = &  F^{p,1}
\end{eqnarray}
and will be detailed in section \ref{subset:dual_potential}.
Note that in the above expression for the dual potential, a prefactor $1/2$ instead of $1/4$ is used for
the crossing-asymmetric part because it does not have the full crossing-symmetries and thus can not be anti-symmetrized; while the crossing-symmetric 
part can be anti-symmetrized and thus carries a prefactor $1/4$. When writing down the above concise
expression for the dual fermion action, we have to impose the constraint that  Hartree-like diagrams in 
the self-energy calculation constructed from $V^{p,0}$ should be eliminated at the one-particle level, 
while at the two-particle level the vertical component of the particle-hole (p-h) two-particle  Green function should be 
canceled by the vacuum term. This constraint stems from taking the replica limit.~\cite{h_terletska_13,m_jarrell_01a,Atland}
Both $V^{p,0}$ and $V^{p,1}$ are for the particle-particle (p-p) channel, and they are related to their counterparts 
for the particle-hole (p-h) channel by the crossing symmetry~\cite{s_yang_09}
\begin{equation}
V^{0}(w,w') = -V^{p,0}(w,w')
\end{equation}
and
\begin{equation}
V^{1}(\nu)_{w,w'} = -V^{p,1}(w+w'+\nu)_{-w',-w}.
\end{equation}

\subsection{Algorithm}

\begin{figure}[tbh]
\centerline{ \includegraphics[clip,scale=0.65]{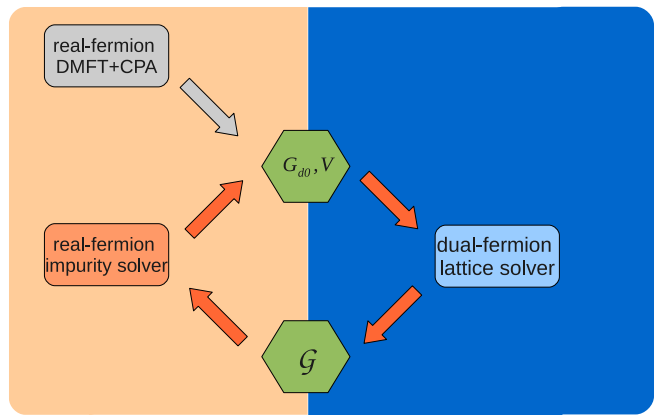}}

\caption{(Color online). Algorithm for the dual fermion approach. The orange region (left half)
is for the real fermion, where the important on-site correlations
are taken into account by numerical exact methods, such as the quantum Monte Carlo (QMC) method. 
The blue region (right half) is for the dual fermion, where the intermediate-length-scale
correlations ignored at the DMFT+CPA calculation are restored systematically.
The connection between these two regions is the dual fermion mapping.}

\label{fig:algorithm} %
\end{figure}
Like the conventional dual fermion algorithm for interacting systems, the dual fermion algorithm for an interacting 
disordered system can be represented schematically by Fig.~\ref{fig:algorithm}. 
We start on the left side from a DMFT+CPA solution of the real fermion system, and then use the information
collected by solving the impurity problem (mainly the one-particle Green function $G_{imp}$,
self-energy $\Sigma_{imp}$, and two-particle
Green function $\chi_{imp}$) to parametrize the dual fermion system in the right half, i.e.,\ construct the bare 
dual fermion Green function $G_{d0}$ and the dual potential $V_d$.
While the local correlations are included in the DMFT+CPA solution, the nonlocal corrections
are incorporated through the dual fermion part, which is calculated using a standard perturbation expansion
in the $V_d$ term. 
After the dual fermion system is solved, we map it back to the real fermion system with the nonlocal 
corrections now included in the lattice self-energy  $\Sigma(w,{\bf k})$ and  Green function $G(w,{\bf k})$.  We then 
solve the impurity problem again starting with an updated impurity-excluded Green function
$\mathcal{G}(w)$. These steps are repeated until self-consistency is 
achieved with 
\begin{equation}
\sum_k G_d(w,{\bf k}) = 0,
\end{equation}
 i.e. with the local contribution to the dual fermion Green function
$G_d(w,{\bf k})$ being zero ~\cite{Rubtsov08}. This condition would fix the arbitrary function $\Delta$ introduced
during the dual-fermion mapping in Appendix \ref{set:df-mapping} and eliminate the first-order contribution
to the self-energy on the dual-fermion lattice.

\subsection{Dual potential}
\label{subset:dual_potential} 

In the algorithm described above, the  non-trivial part is the measurement of the two-particle Green functions.
This is due to the requirement that the crossing-asymmetric and crossing-symmetric contributions must be separated and treated differently. 
We here propose a two-step procedure to measure the two-particle Green functions:
\begin{itemize}
\item quantum averaging (integrate out the  Coulomb term)
\item disorder averaging (integrate out the  disorder term)
\end{itemize}
By this procedure, the crossing-asymmetric and crossing-symmetric components can be separated 
as detailed  in the following. 

\begin{figure}[th]
\centerline{ \includegraphics[clip,scale=0.55]{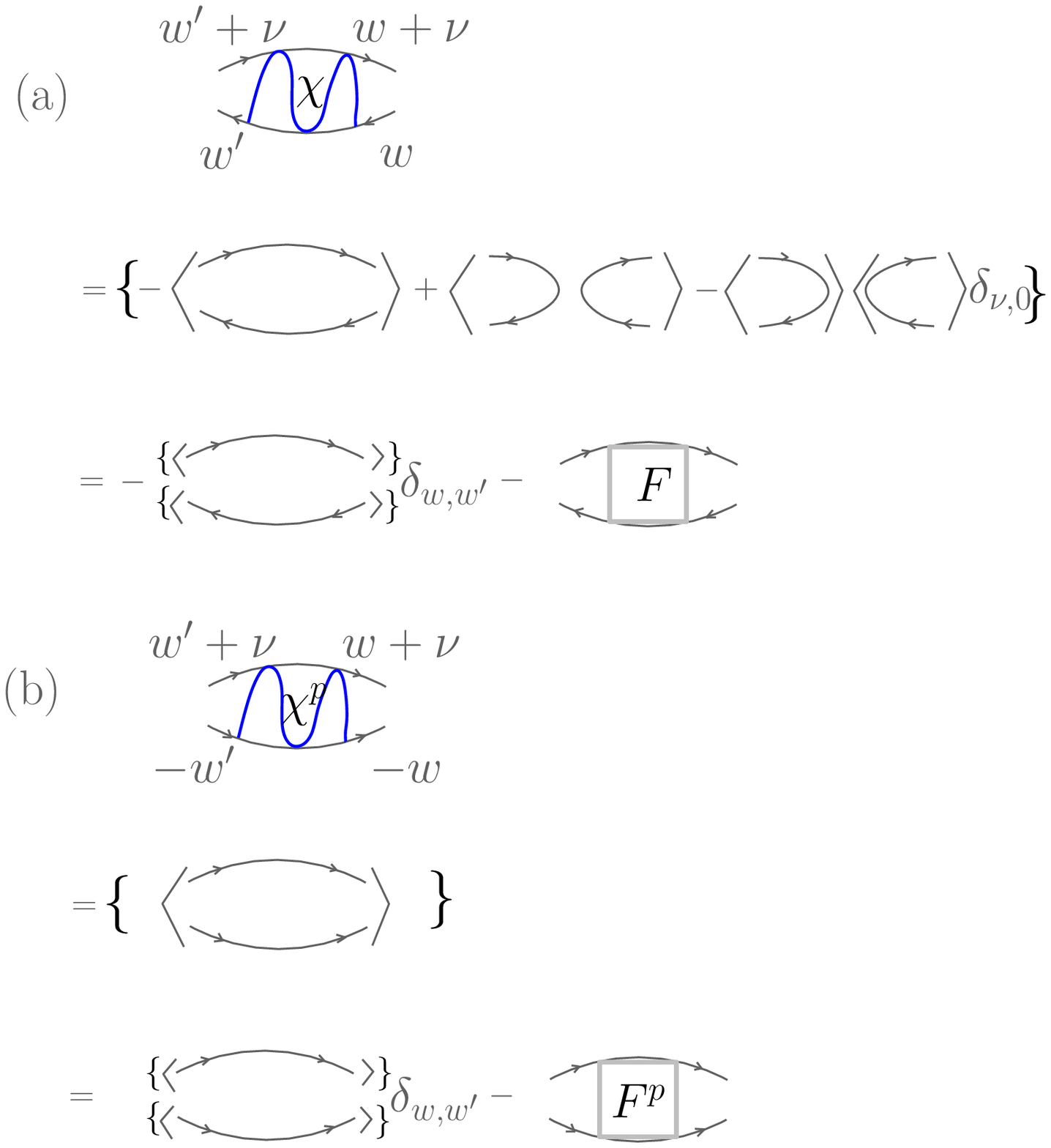}} 
\caption{(Color online). Measurement formula for the two-particle Green functions
and the defining equation for the full vertex in the particle-hole (p-h) channel (a) and particle-particle 
(p-p, $X^p$ denoting a quantity in this channel) channel (b). 
Note that the angle brackets represent the quantum averaging 
while the curly brackets represent the disorder averaging.}
\label{fig:chi_ph} 
\end{figure}
In Fig. \ref{fig:chi_ph} (a) we show for the case of the p-h channel that the full two-particle Green function $\chi$
can be measured as 
(spin indices are suppressed to simplify the expressions, and hereafter, $g$ represents the one-particle Green function 
for a given decoupling field (for quantum averaging) and disorder configuration (for disorder averaging), 
while $G$ and $\chi$ are the fully dressed one-particle and
two-particle Green functions respectively)
\begin{eqnarray}
\chi(\nu)_{w,w'} &=& \{-<g(w+\nu,w'+\nu)g(w',w)> \nonumber \\
                       && +<g(w+\nu,w)g(w',w'+\nu)> \nonumber \\
                       && -<g(w,w)><g(w',w')>\delta_{\nu,0}\},
\end{eqnarray}
in which the angle brackets represent the quantum averaging while the curly brackets represent the disorder averaging.
The full vertex is defined according to the following equation
\begin{eqnarray}
& & \chi(\nu)_{w,w^{\prime}} =   -G(w+\nu)G(w)\delta_{w,w^{\prime}}\nonumber \\
                       &   & -TG(w+\nu)G(w)F(\nu)_{w,w^{\prime}}G(w^{\prime}+\nu)G(w^{\prime}).
\end{eqnarray}
Similarly, as shown in Fig. \ref{fig:chi_ph} (b), we have
\begin{eqnarray}
\chi^p(\nu)_{w,w'} &=& \{<g(w+\nu,w'+\nu)g(-w,-w')>  \}\nonumber\\
& =&    G(w+\nu)G(-w)\delta_{w,w^{\prime}}\, -TG(w+\nu)G(-w) \nonumber \\
                       &   & \times F^p(\nu)_{w,w^{\prime}}G(w^{\prime}+\nu)G(-w^{\prime})
\end{eqnarray}
for the p-p channel.

\begin{figure}[th]
\centerline{ \includegraphics[clip,scale=0.6]{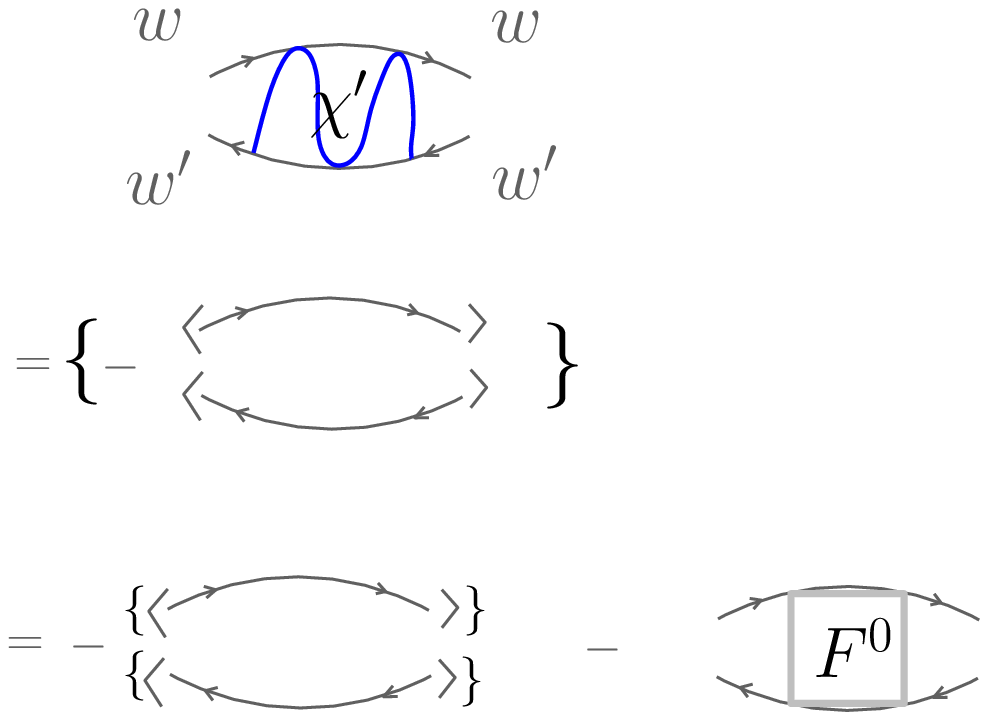}} 
\caption{(Color online). Measurement formula for the crossing-asymmetric component of the two-particle Green function in the p-h
channel and the defining equation for the full vertex. }
\label{fig:chi_ph_0} 
\end{figure}

Note that the two-particle Green function measured in this manner still contains both the crossing-asymmetric and crossing-symmetric contributions. 
However, by analyzing the  diagrams contributing to each component, we realize that the crossing-asymmetric component 
can be measured individually as
(see Fig. \ref{fig:chi_ph_0} for the Feynman diagram)
\begin{eqnarray}
\chi^\prime(w,w') &=& \{ -<g(w,w)><g(w',w')>\}.
\end{eqnarray}
The full vertex for the crossing-asymmetric component can then again be calculated using the definition equation
\begin{eqnarray}
& & \chi^\prime(w,w')  =   -G(w)G(w')\nonumber \\
                       &   & -TG(w)G(w')F^0(w,w')G(w)G(w^{\prime}).
\end{eqnarray}

\begin{figure}[tbh]
\centerline{ \includegraphics[clip,scale=0.45]{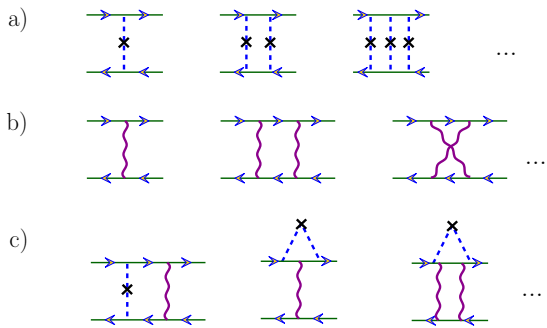}}
\caption{(Color online). Feynman diagrams contributing to the full vertex in the p-h channel.
Group a) are  the diagrams due to disorder contributions only.
Note between the two fermion lines only  disorder scattering processes appear
(dashed line and black crosses). Group b) diagrams are from the interactions
with only Coulomb interaction lines (wavy lines). And 
group c) displays the mixed contributions.}
\label{fig:F} %
\end{figure}

\begin{figure}[tbh]
\centerline{ \includegraphics[clip,scale=0.6]{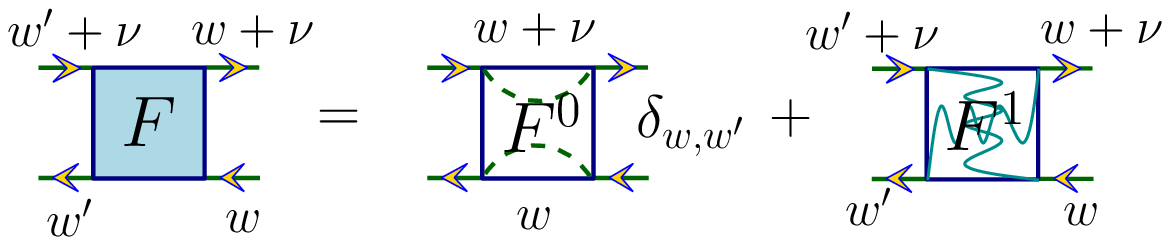}} 
\caption{(Color online). Decomposing the full vertex into crossing-asymmetric and crossing-symmetric scattering
processes. The crossing-asymmetric component only depends on two frequencies and is coming from the 
disorder scattering at the two-particle level, while the crossing-symmetric component is from
the scattering due to Coulomb interaction or from the combined interplay
 of Coulomb interaction and disorder scattering. }
\label{fig:F-decomp} %
\end{figure}

Let us now turn to the calculation of the
crossing-symmetric component $F^1(\nu)_{w,w'}$. 
%
%
Diagrams for the full vertex are illustrated in Fig.~\ref{fig:F}, where the scattering 
from disorder is represented by a dashed line and black crosses, and the interaction 
by a wavy line.  One obtains pure disorder diagrams (where only disorder scatterings 
appear in the connection of two fermion lines, only two frequencies are needed and thus
can be expressed as $X(w,w')$), pure interacting diagrams (only 
interaction scatterings in the connections) and the mixed diagrams:
\begin{equation}
F(\nu)_{w,w'} = F_{D}(w+\nu, w)\delta_{w,w'} + F_{U}(\nu)_{w,w'} + F_{mix}(\nu)_{w,w'}.
\end{equation}
To be able to do the calculation in the dual fermion formalism, one has to separate 
out the pure disorder contributions $F_{D}$. Then the full vertex can be divided into two 
parts as shown in Fig.~\ref{fig:F-decomp}
\begin{equation}
F(\nu)_{w,w'} = F^{0}(w+\nu, w)\delta_{w,w'} + F^{1}(\nu)_{w,w'}
\label{eq:decomp} 
\end{equation}
with
\begin{equation}
F^{0}(w,w')\equiv F_{D}(w,w')
\end{equation}
and finally
\begin{eqnarray}
F^{1}(\nu)_{w,w'} &\equiv& F_{U}(\nu)_{w,w'}+F_{mix}(\nu)_{w,w'} \nonumber \\
                &=&      F(\nu)_{w,w'}-F^{0}(w+\nu, w')\delta_{w,w'} 
\end{eqnarray}
Note that while $F^{1}$ is fully crossing-symmetric and thus can
be treated like a conventional vertex function in a clean interacting
system, $F^{0}$ does not have full crossing symmetry. Specifically,
each crossing symmetry involving the particle-hole vertical channel
vertex is broken because that single channel contribution is absent
from the construction.

\section{Results}

To test our generalization of the dual fermion approach, we apply it to 
the simplest interacting disordered fermion system, the 
Anderson-Falicov-Kimball model
\begin{equation} 
\mathcal{H} = \sum_{{\bf k}} (\epsilon_k-\mu) c_{{\bf k}}^{\dagger}c_{{\bf k}} 
            - \sum_{i}v_{i}n_{i}^{c}+U\sum_{i}n_{i}^{c}n_{i}^{f},
\end{equation}
where $\epsilon_k$ is the dispersion for the itinerant c-electrons (only the nearest neighbor hopping is included), the on-site 
disorder potential $v_i$ follows the ``box" probability distribution of Eq.~(\ref{eq:distribution})
and U is the on-site Coulomb interaction between itinerant c-electrons and immobile f-electrons.
In the limit of no Coulomb interaction, this model reduces to the Anderson disorder model and 
the dual-fermion calculation is presented in our previous contribution~\cite{h_terletska_13}.
In the other limit of no disorder, it reduces to the Falicov-Kimball model 
for which both Quantum Monte Carlo and DCA results are available in the literature~\cite{Maska_06,Hettler98,Hettler00}, 
and the dual fermion method is also applied on this model recently~\cite{df_review, Antipov_13}.

In Sections \ref{sec:gl} and \ref{sec:dg} we will first look into corrections from the DF to the DMFT+CPA results on the 
one-dimensional (1D) lattice using the self-consistent second-order method
to solve the DF lattice problem (DF-2nd) as an example (see Appendix \ref{sec:DF-2nd} for details). 
In Section \ref{sec:phase-diagram} we will carry out a detailed survey of the U-D phase-diagram for the two-dimensional (2D) 
lattice  at a fixed temperature $T=0.05$ (with $4t=1$) using DF-2nd and the fluctuation-exchange (FLEX) approximation 
to solve the DF lattice problem (DF-FLEX), see Appendix \ref{sec:DF-FLEX} for additional details on the DF-FLEX approach.
The filling is fixed at half-filling for both c- and f-electrons.

\subsection{Local Green function}
\label{sec:gl} %
\begin{figure}[tbh]
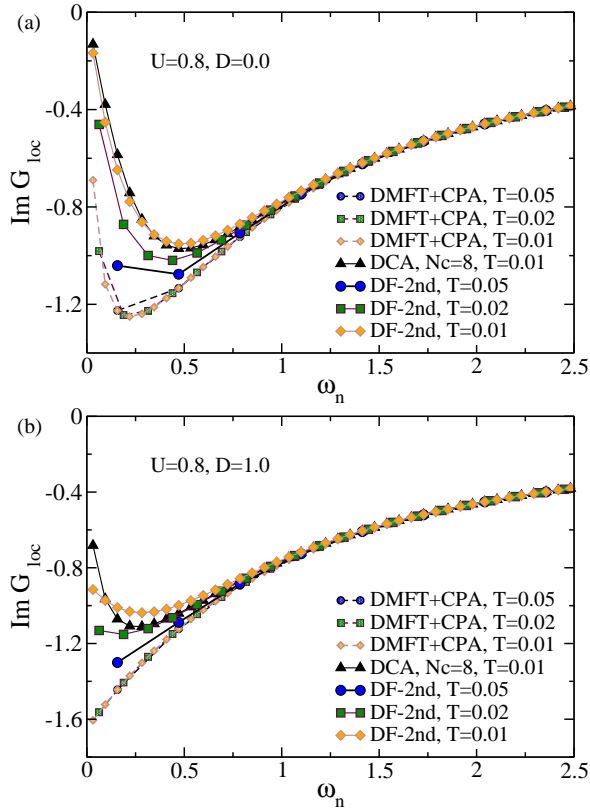

\centerline{ \includegraphics[clip,scale=0.3]{./gl_U08_V0.eps}}

\centerline{ \includegraphics[clip,scale=0.3]{./gl_U08_V1.eps}}

\caption{(Color online). Comparison of the local Green function calculated by the DMFT+CPA 
and DF approaches for the 1D lattice. Note that the former is essentially temperature independent,
while the latter is temperature-dependent and shows very different
behavior which is consistent with DCA results (black up-triangles).}

\label{fig:gl} %
\end{figure}
Figure \ref{fig:gl} shows the Matsubara frequency dependence of the local Green function 
calculated from both DMFT+CPA and DF approaches for the 1D lattice 
at $U=0.8$ for zero $D=0$ and finite $D=1.0$ disorder strength 
at different temperatures $T=0.01, 0.02, 0.05$.
For clean system at U=0.8 and D=0 (Fig.\ref{fig:gl} (a)), 
both methods show an insulator-like behavior,
which can be inferred from  the  imaginary part of the local Green function converging to zero as function of temperature
for the lowest Matsubara frequency. However, while the DMFT+CPA results
are temperature independent due to the neglect of the non-local correlations, 
there is a significant temperature dependence in the results from the DF approach. Moreover,
they appear to be consistent with the DCA calculation for a cluster of size $Nc=8$ at $T=0.01$ included in 
Fig.\ \ref{fig:gl}  as filled black up-triangles.

Similarly, for finite disorder with $D=1.0$ (Fig. \ref{fig:gl} (b)), the DF results recover the important temperature
dependence from the non-local correlations, which are again absent in the DMFT+CPA calculation. 
Due to the disorder, the system becomes less insulating 
as compared to the clean case $D=0$  (contrast the lowest Matsubara frequency results in Fig. \ref{fig:gl}$(a)$ and $(b)$).
This is well-captured by the DF calculation
which is also consistent with the DCA results, while
the DMFT+CPA approach strongly over-estimates this effect.

\subsection{Correction from the dual fermion calculation}
\label{sec:dg} %

\begin{figure}[tbh]
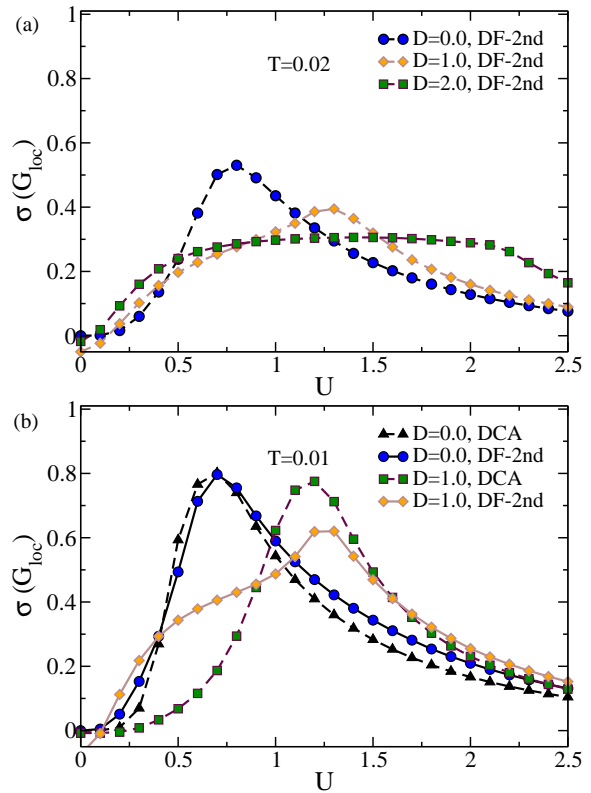

\centerline{ \includegraphics[clip,scale=0.3]{./gl_diff_UV_T002.eps}}

\centerline{ \includegraphics[clip,scale=0.3]{./gl_diff_UV_T001.eps}}

\caption{(Color online). Relative correction from the dual fermion approach
to the local Green function at the lowest Matsubara frequency ($iw=i\pi T$)
for various parameters of the 1D lattice. The corrections
are minimized for both weak- and large-U limits and maximized for values of U around
the band-with. The peak position shifts to larger U with
increasing disorder strength. When decreasing the temperature,
the corrections all increase. This behavior is consistent with DCA results.}

\label{fig:gl-diff} %
\end{figure}
In order to quantify how strong the corrections due to the non-local correlations from the DF 
approach are, we introduce the following quantity
\begin{equation} 
\sigma (G_{loc}) \equiv 
\frac{Im G^{DF}_{loc}(i\pi T)- Im G^{DMFT+CPA}_{loc}(i\pi T)}{|Im G^{DMFT+CPA}_{loc}(i\pi T)|}, 
\end{equation}
which represents the relative difference of the imaginary part of the local Green function
at the lowest Matsubara frequency. Results for the 1D lattice are shown in Fig. \ref{fig:gl-diff}. 

For both temperatures, T=0.02 (Fig. \ref{fig:gl-diff}$(a)$) and T=0.01 (Fig. \ref{fig:gl-diff}$(b)$), the corrections are weak
 in both small and 
large $U$ limits, reaching their maximum around $U\approx W=1$, $W$ being the bandwidth. This verifies our belief that
DMFT+CPA for the one-particle Green's function works best for $U\ll W$ and $U\gg W$, 
while for $U\approx W$
the kinetic and interaction parts of the Hamiltonian are strongly competing and non-local correlations
become more important. Thus we will naturally expect an enhanced correction
from the DF calculation in this region. The corrections are around 30 percent for T=0.02, and they 
increase to around 60 percent when the temperature is decreased to T=0.01. One interesting
observation is that the peak in the plot shifts to
larger U values and gradually smooths out when the disorder is increased.

\subsection{U-D phase diagram for the two-dimensional lattice}
\label{sec:phase-diagram} %

Let us now turn to the phase diagram of the $2D$ Anderson-Falicov-Kimball model.

\begin{figure}[tbh]
\centerline{ \includegraphics[clip,scale=0.48]{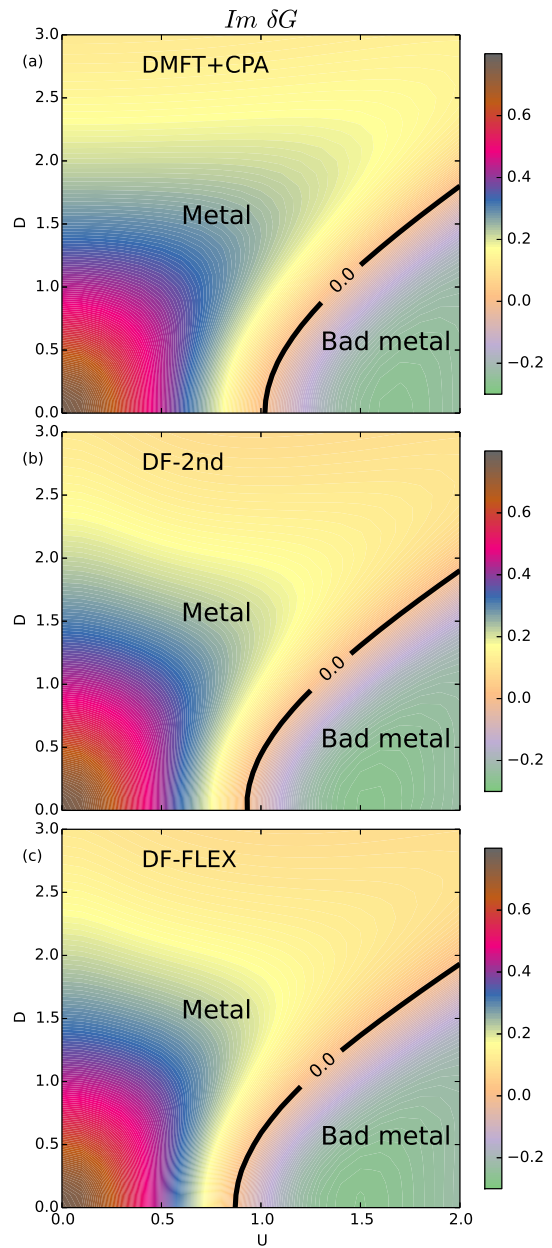}}
\caption{(Color online). U-D  phase diagram of the $2D$  Anderson-Falicov-Kimball model determined from 
the difference between the imaginary part of the local Green function at the two lowest Matsubara frequencies.
Panel (a) displays the DMFT+CPA, (b) the DF-2nd and (c) the DF-FLEX results, respectively, at $T=0.05$ ($4t=1$). 
The metal-bad-metal crossover is indicated by the black solid line, 
while the Anderson localization transition cannot be detected by the one-particle
Green function.}
\label{fig:dg_UV} %
\end{figure}

An easy way to analyze the metal-insulator transition due to
crossing-symmetric scattering processes from the Coulomb interaction is by
looking at the difference between the imaginary part of the local Green function at the two lowest Matsubara frequencies:
\begin{equation}
Im\; \delta G = Im G_{loc}(3i\pi T) - Im G_{loc}(i\pi T).
\end{equation}
When decreasing the temperature, the imaginary part of local Green function converges 
to zero for the insulating phase, while it diverges for the metallic phase. This different 
behavior can be captured with the above  quantity, which is 
negative for the insulating phase, and positive for the metallic phase. 
Note that this distinction between metal and insulator phases is accurate only at zero temperature. 
For finite temperature, the insulator determined in this way could be just a bad metal in reality.
So in the following, we will call it bad metal phase. The transition is actually a crossover indicated 
by the changing of the sign of $Im \, \delta G$. Fig. \ref{fig:dg_UV} shows the U-D phase diagram thus determined.

Both disorder and interaction tend to decrease the mobility of itinerant particles. However,
in  Fig. \ref{fig:dg_UV} only interaction drive the system to an insulating phase, while it remains
metallic even for large disorder strength. This is due to the arithmetic averaging nature of 
the one-particle Green function within the DF approach,
which does not distinguish between extended and localized electrons.
The positive slope of the metal-bad-metal crossover line indicates that
the disorder scattering helps particle mobility.
Including more ladder-type diagrams, for example within a DF-FLEX calculation (Fig. \ref{fig:dg_UV}$(b)$),
tends to move the crossover line to a smaller value of the interaction, and this shifting is less pronounced for larger 
disorder strengths.

\begin{figure}[tbh]
\centerline{ \includegraphics[clip,scale=0.45]{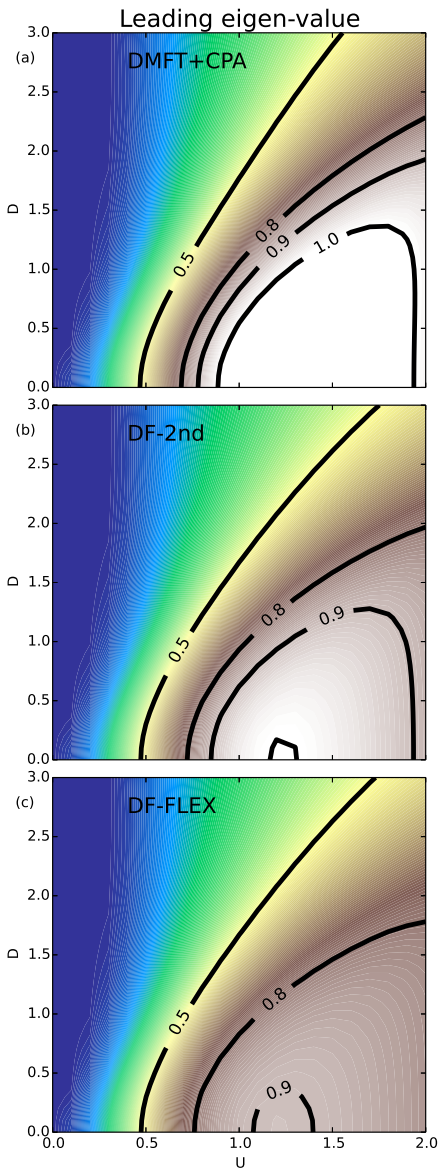}}
\caption{(Color online). U-D phase diagram of the $2D$ Anderson-Falicov-Kimball model determined from the leading-eigen-value (LEV)
for the charge density wave (CDW) channel calculated by DMFT+CPA (a), DF-2nd (b) and DF-FLEX (c) at $T=0.05$ ($4t=1$).  
The closer of the LEV to one, the more susceptible is the system to the CDW pairing
formation. The region most susceptible to CDW ordering resembles that of the 
bad metal region determined in Fig. \ref{fig:dg_UV}, suggesting that the metal-bad-metal
crossover is driven by CDW ordering.}
\label{fig:lev_UV} %
\end{figure}

To analyze the driving force of the metal-bad-metal crossover detected by the difference between the imaginary part 
of the two lower Matsubara frequencies local Green functions, we show  in Fig. \ref{fig:lev_UV} the phase-diagram determined
by the leading eigen-values (LEV) for the charge-density-wave (CDW) channel. The LEV $\lambda$ is calculated
by solving the eigen problem $\Gamma \chi_0 \phi = \lambda \phi$, where $\Gamma$ is the irreducible vertex
and $\chi_0$ is the bare lattice susceptibility. Note that the 
closer the  LEV is to unity, the more susceptible the system becomes to the CDW ordering. We can 
readily observe that the large-LEV region resembles the bad metal region
in Fig. \ref{fig:dg_UV}. This indicates that the metal-bad-metal crossover is driven by
 CDW correlations.

\begin{figure}[tbh]
\centerline{ \includegraphics[clip,scale=0.45]{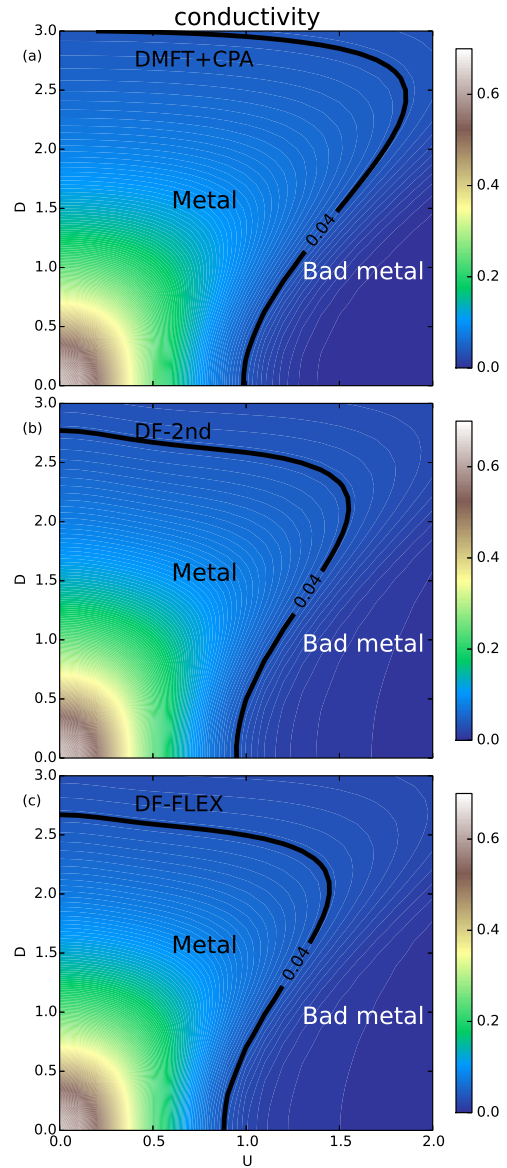}}
\caption{(Color online). U-D phase diagram determined from the conductivity calculated
by DMFT+CPA (a), DF-2nd (b) and DF-FLEX (c) at $T=0.05$. The metal-bad-metal crossover either due to the
Anderson localization or the Coulomb interaction is indicated by the black solid line
with finite but small conductivity $\sigma_{dc}= 0.04$.}
\label{fig:cond_UV} %
\end{figure}
The phase-diagram determined from either $Im \, \delta G$ or the LEV does not provide any signature
of the Anderson localization. To cure such a deficiency, we turn to  
the dc conductivity (see Appendix \ref{set:vertex_manipulation} for the details of its calculation). 
In marked difference with the DMFT+CPA approach, the full vertex correction to the conductivity can be
taken into account using the parquet equation.~\cite{h_terletska_13,s_yang_09}
Results for this quantity are collected in Fig. \ref{fig:cond_UV}, where we present  the phase-diagram as 
determined by the conductivity.

In our calculation, the conductivity remains finite for all values of $D$ and $U$ 
on the 2D lattice. 
Therefore we use a small but finite 
conductivity value $\sigma_{dc}= 0.04$ to delineate metallic and bad metal regions. This value 
is determined by matching the critical U for zero disorder strength to the one 
determined by $Im \, \delta G$, since both approaches should produce consistent result for the clean system. 
With this convention, we observe in Fig. \ref{fig:cond_UV} that the Anderson localization line
connects continuously  to the CDW metal-bad-metal crossover line. 
This is qualitatively similar to the ground-state phase-diagram obtained 
from the typical medium theory for the Bethe lattice ~\cite{Byczuk-phase_diagram,Gusmano-phase_diagram}. The difference is that
 the slope of the crossover line for small U is negative 
which indicates that the Coulomb interaction helps in localizing the particles.

\section{Conclusion}
We have generalized the recently proposed DF approach to treat 
both  disorder and Coulomb interactions. This generalization is possible 
due to the clear separation between the crossing-asymmetric component, due to the disorder scattering, of the 
two-particle level vertex, and the crossing-symmetric component, due to the Coulomb interaction processes
and the combined scattering processes from both disorder and Coulomb interactions. 
Thus the two competing factors can be treated on equal footing in this method. 

We would like to emphasize that the separation and different treatments of
crossing-asymmetric and crossing-symmetric scattering processes are generic for 
two-particle field-theory calculation of interacting disordered systems. Its
application is not limited to the DF method proposed here, and instead
might also find its use in other multi-scale diagrammatic methods, such as
multi-scale many-boby approach~\cite{c_slezak_06b} and dynamical vertex approximation~\cite{Toschi_07}.

To demonstrate
the algorithm, we apply the method to the $1D$ Anderson-Falicov-Kimball at half filling and compare
our results to those obtained with other established approximations, viz the 
DMFT-CPA  and larger cluster DCA calculations. We observe that our approach gives satisfying
results, significantly improving on the DMFT-CPA and systematically approaching the DCA simulations. 

As an important and challenging application we study the phase diagram of the 2D Anderson-Falicov-Kimball system 
using both single- and two-particle quantities. 
We observe that the interaction-driven metal bad-metal crossover is clearly due to
CDW correlations. Increasing the disorder, on the other hand, does not seem to lead to a metal-bad-metal crossover based on
the behavior of the one-particle properties. Since for the $2D$ model one however should observe signs of Anderson localization, 
we resort to a quantity that should show this effect. We here use the conductivity, which indeed gives a crossover to a 
bad metal with increasing disorder strength which competes with the Coulomb interaction. This latter
competition leads to an interesting re-entrance behavior of the metal-bad-metal crossover line for small to moderate
disorder strength, increasing the stability of the metallic phase. 

These results indeed show that the algorithm introduced here is capable of treating interactions and
disorder on the same footing, with results that significantly go beyond standard DMFT/CPA calculations.
 Further applications to the full Anderson-Hubbard model are on their way.

\begin{acknowledgments}

This work is supported by DOE SciDAC grant DE-FC02-10ER25916 (SY and MJ) and BES CMCSN 
grant DE-AC02-98CH10886 (HT).  
Additional support was provided by NSF EPSCoR Cooperative Agreement No. EPS-1003897 (SY, HT, ZY), 
NSF OISE-0952300 (SY, JM),
and by DFG through research unit FOR 1807 (TP and PH). Computer support is provided 
by the NSF Extreme Science and Engineering Discovery Environment (XSEDE) under grant number DMR100007, the
Louisiana Optical Network Initiative, and HPC@LSU computing, and  
by the Gesellschaft f\"ur wissenschaftliche Datenverarbeitung G\"ottingen (GWDG) and the GoeGrid project.
\end{acknowledgments}

\appendix
 
\section{Dual fermion mapping}
\label{set:df-mapping} 

In this section, we will derive the dual fermion formalism using the replica technique.
We consider the Anderson-Hubbard model described by Eq. \ref{eq:AH}.

The disorder averaged lattice Green function is given by 
\begin{equation}
G_\sigma(w,{\bf k}) = -\frac{\delta}{\delta\eta_{w{\bf k}\sigma}}\left\{\ln Z^{v}[\eta_{\omega{\bf k}\sigma}]\right\}|_{\eta_{w{\bf k}\sigma}=0},
\label{GF-defintion}
\end{equation}
 with $\left\{(...)\right\}=\int dvp(v)(...)$ indicating a disorder
averaged quantity, $X^v$ representing the quantity $X$ is disorder configuration dependent and $\eta_{w{\bf k}\sigma}$ being a source field. 
The partition function for a given disorder configuration $\{v_i\}$ is defined as
\begin{equation}
Z^{v}[\eta_{w{\bf k}\sigma}] = \int D\bar{c}Dce^{-S^{v}[\eta_{w{\bf k}\sigma}]},
\end{equation}
where $Dc\equiv\prod_{w{\bf k}\sigma}{dc_{w{\bf k}\sigma}}$, and the action is itself defined as
\begin{eqnarray}
& & S^{v}[\eta_{w{\bf k}\sigma}] = \sum_{w{\bf k}\sigma}\bar{c}_{w{\bf k}\sigma}(-iw+\varepsilon_{{\bf k}}-\mu+\eta_{w{\bf k}\sigma})c_{w{\bf k}\sigma} \nonumber \\
&+&\sum_{i\sigma}v_{i} \int_0^\beta d\tau n_{i\sigma}(\tau)
 + U\sum_{i} \int_0^\beta d\tau n_{i\uparrow}(\tau)n_{i\downarrow}(\tau),\nonumber \\
\label{eq:action}
\end{eqnarray}
 where $iw=i(2n+1)\pi T$ are the Matsubara frequencies, $\varepsilon_{{\bf k}}$
is the lattice bare dispersion, $\mu$ is the chemical potential, and $U$ the Coulomb interaction.
In the following, the explicit dependence on spin index $\sigma$ and 
source term $\eta_{w{\bf k}\sigma}$ will be hidden
to simplify the expressions. Using the replica trick 
\begin{equation}
\ln Z = \lim_{m\rightarrow0} \frac{Z^{m}-1}{m}, 
\end{equation}
where $m$ replicas are introduced,
we can express the disorder-averaged Green function as 
\begin{equation}
G(w,{\bf k})=-\lim_{m\rightarrow0}\frac{1}{m}\frac{\delta}{\delta\eta_{w{\bf k}}}
\left\{
\int\mathcal{D}\bar{c}\mathcal{D}ce^{-S^{v_i}[c^{\alpha},\bar{c}^{\alpha}]}
\right\}|_{_{\eta_{w{\bf k}}=0}},
\label{app-eq:GF}
\end{equation}
where $\mathcal{D}c\equiv\prod_{w{\bf k}\alpha}{dc_{w{\bf k}}^{\alpha}}$, 
and $\alpha$ is the replica index. The replicated lattice action  is
\begin{eqnarray}
S^{v_i}[c^{\alpha},\bar{c}^{\alpha}]
  &=& \sum_{w{\bf k}\alpha}\bar{c}_{w{\bf k}}^{\alpha}(-iw+\varepsilon_{{\bf k}}-\mu+\eta_{w{\bf k}})c_{w{\bf k}}^{\alpha} \nonumber \\
  &+&  \sum_{i\alpha}v_{i}\int_{0}^{\beta}d\tau n_{i}^{\alpha}(\tau)  \nonumber \\
  &+& U\sum_{i\alpha}\int_{0}^{\beta}d\tau n_{i\uparrow}^{\alpha}(\tau) n_{i\downarrow}^{\alpha}(\tau).
\label{app-eq:action_original}
\end{eqnarray}
The disorder averaging can be formally done, and thus we obtain 
\begin{eqnarray}
S[c^{\alpha},\bar{c}^{\alpha}]
  &=& \sum_{w{\bf k}\alpha}\bar{c}_{w{\bf k}}^{\alpha}(-iw+\varepsilon_{{\bf k}}-\mu+\eta_{w{\bf k}})c_{w{\bf k}}^{\alpha} \nonumber \\ 
  &+&  \sum_{i}W(\tilde{n}_{i}) 
   + U\sum_{i\alpha}\int_{0}^{\beta}d\tau n_{i\uparrow}^{\alpha}(\tau) n_{i\downarrow}^{\alpha}(\tau). \nonumber \\ 
\label{app-eq:action_av}
\end{eqnarray}
Note that the Coulomb interaction term remains the same form, and a new elastic, effective interaction between
electrons of different replicas $W(\tilde{n}_{i})$ appears due to the disorder scattering. 
The latter is local in space and non-local in time, and could be expressed through local cumulants $<v_i^{l}>_{c}$ as \cite{m_jarrell_01a}
\begin{eqnarray}
e^{-W(\tilde{n}_{i})} & = & \int dv_{i}p(v_{i})e^{-v_{i}\sum_{\alpha}\int d\tau n_{i}^{\alpha}(\tau)}\nonumber \\
 & = & e^{-\sum_{l=2}^{\infty}\frac{1}{l!}<v_i^{l}>_{c}\left(\sum_{\alpha}\int d\tau n_{i}^{\alpha}(\tau)\right)^{l}}.
\label{eq:cumulant}
\end{eqnarray}

Similarly to the non-interacting disorder fermionic systems~\cite{h_terletska_13}, we follow four steps
to derive the DF formalism for the interacting disorder models.
First, we introduce an effective single-site impurity reference problem
by formally rewriting the original action as 
\begin{equation}
S=\sum_{i}S_{imp}[\bar{c}_i^{\alpha},c_i^{\alpha}]-\sum_{w{\bf k}\alpha}{\bar{c}_{w{\bf k}}^{\alpha}(\Delta_{w}-\varepsilon_{{\bf k}}-\eta_{w{\bf k}})c_{w{\bf k}}^{\alpha}},
\label{eq:action_with_imp}
\end{equation}
 with an effective impurity action (containing both the Coulomb and disorder interactions,
$W(\tilde{n}_{i}$)) 
\begin{eqnarray}
S_{imp} &=& \sum_{\alpha w}\bar{c}_{wi}^{\alpha}(-iw-\mu+\Delta_{w})c_{wi}^{\alpha}\nonumber \\
 &+& W(\tilde{n}_{i}) + U\sum_{\alpha}\int_{0}^{\beta}d\tau n_{i\uparrow}^{\alpha}(\tau) n_{i\downarrow}^{\alpha}(\tau).
\end{eqnarray}
 Here $\Delta_{w}$ is a local, and yet unknown, hybridization function
describing the interaction of the impurity with the effective medium.
As in the original DF formalism, it is assumed
that all the properties of the impurity problem, i.e., the one-particle
Green function \begin{equation}
G_{imp}(w)=-\lim_{m\rightarrow0}\frac{1}{m}\sum_{\alpha=1}^{m}{\displaystyle \int\mathcal{D}\bar{c}\mathcal{D}c\,
 c_{w}^{\alpha}\bar{c}_{w}^{\alpha}e^{-S_{imp}},}
\end{equation}
 and the two-particle Green functions which contain effects from both Coulomb
interaction and disorder
\begin{eqnarray}
&& \chi_{imp}^p(\nu)_{w,w'} \nonumber \\
&=& \lim_{m\rightarrow0}\frac{1}{m}
\sum_{\alpha,\beta=1}^{m}{\displaystyle \int\mathcal{D}\bar{c}\mathcal{D}c \,
c_{w+\nu}^{\alpha}c_{-w}^{\beta}\bar{c}_{-w'}^{\alpha}\bar{c}_{w'+\nu}^{\beta}\, e^{-S_{imp}}} 
\nonumber \\
\end{eqnarray}
 can be calculated. See \ref{subset:dual_potential} in the main text about how to 
measure them in the real calculation.
These Green functions are local quantities. Our
task is to express the original lattice Green function and other properties
via quantities of the DMFT+CPA impurity problem. What has been accomplished
so far in Eq.~$(\ref{eq:action_with_imp})$ is that the local part
of the lattice action has been moved to the effective impurity.

In the second step of the DF procedure we introduce auxiliary
({}``dual'' fermions) degrees of freedom. In doing so, we transfer
the non-local part of the action in Eq.~$(\ref{eq:action_with_imp})$
to the dual variables. As a result, the original real fermions carry
information about the local part only. The transformation to dual
fermions is done via a Gaussian transformation of the non-local part
of Eq.~$(\ref{eq:action_with_imp})$,

\begin{eqnarray}
&& e^{\bar{c}_{w{\bf k}}^{\alpha}A_{w{\bf k}}^{2}c_{w{\bf k}}^{\alpha}} \nonumber\\
&=&\frac{A_{w{\bf k}}^{2}}{\lambda_{w}^{2}}\int\mathcal{D}\bar{f}\mathcal{D}f
e^{-\lambda_{w}(\bar{c}_{w{\bf k}}^{\alpha}f_{w{\bf k}}^{\alpha}+\bar{f}_{w{\bf k}}^{\alpha}c_{w{\bf k}}^{\alpha})
-\frac{\lambda_{w}^{2}}{A_{w{\bf k}}^{2}}\bar{f}_{w{\bf k}}^{\alpha}f_{w{\bf k}}^{\alpha}},\label{eq:Hub-Strat}\nonumber\\
\end{eqnarray}
 with $A_{w{\bf k}}^{2}=(\Delta_{w}-\varepsilon_{{\bf k}}-\eta_{w{\bf k}})$, and $\lambda_{w}$
yet to be specified.

With such a transformation, the lattice Green function of Eq.~$(\ref{app-eq:GF})$
can be rewritten as \begin{eqnarray}
G(w,{\bf k}) & = & -\lim_{m\rightarrow0}\frac{1}{m}\frac{\delta}{\delta\eta_{w{\bf k}}}\frac{\left(\Delta_{w}-\varepsilon_{{\bf k}}-\eta_{w{\bf k}}\right)}{\lambda_{w}^{2}}\nonumber \\
 & \times & \int\mathcal{D}\bar{f}\mathcal{D}f\, e^{-\sum_{w{\bf k}\alpha}\lambda_{w}^{2}\bar{f}_{w{\bf k}}^{\alpha}\left(\Delta_{w}-\varepsilon_{{\bf k}}-\eta_{w{\bf k}}\right)^{-1}f_{w{\bf k}}^{\alpha}}\nonumber \\
 & \times & \int\mathcal{D}\bar{c}\mathcal{D}c\, e^{-\sum_{i}S_{site}^{i}[\bar{c}_{i}^{\alpha},c_{i}^{\alpha};\bar{f}_{i}^{\alpha},f_{i}^{\alpha}]}|_{_{\eta_{w{\bf k}}=0}},\nonumber \\
\label{GF_with_S_site}\end{eqnarray}
in which the replicated action for site $i$ is of the form \begin{equation}
S_{site}^{i}=S_{imp}+\sum_{\alpha w}\lambda_{w}\left(\bar{c}_{iw}^{\alpha}f_{iw}^{\alpha}+\bar{f}_{iw}^{\alpha}c_{iw}^{\alpha}\right).\label{S_site}\end{equation}
 In Eq.~$(\ref{GF_with_S_site})$ the inter-site coupling is transferred
to a coupling between dual fermions. 

In the third step of the DF mapping, we integrate out the real fermions
from the local site action $S_{site}^{i}$ for each site $i$ separately,
i.e., \begin{eqnarray}
 &  & \int\prod_{\alpha w}d\bar{c}_{i}^{\alpha}dc_{i}^{\alpha}e^{-S_{site}[\bar{c}_{i}^{\alpha},c_{i}^{\alpha};\bar{f}_{i}^{\alpha},f_{i}^{\alpha}]}\nonumber \\
 & = & Z_{imp}e^{-\sum_{w\alpha}\lambda_{w}^{2}G_{imp}(w)\bar{f}_{iw}^{\alpha}f_{iw}^{\alpha}-V_{d,i}^{\alpha,\beta}[\bar{f}_{i}^{\alpha},f_{i}^{\beta}]},\label{Vdf_def}\end{eqnarray}
 in which $Z_{imp}$ is the partition function for the replicated
impurity system

\begin{equation}
Z_{imp}=\int\prod_{\alpha w}d\bar{c}_{i}^{\alpha}dc_{i}^{\alpha}e^{-S_{imp}[\bar{c}_{i}^{\alpha},c_{i}^{\alpha}]}.
\end{equation}
As in the clean case,
formally this can be done up to infinite order, which makes the mapping
to the DF variables exact. Choosing for convenience $\lambda_{w}=G_{imp}^{-1}(w)$,
the lowest-order of the replicated DF potential $V_{d,i}^{\alpha,\beta}[\bar{f}_{i}^{\alpha},f_{i}^{\beta}]$
 reads as 
\begin{eqnarray}
V_{d,i}^{\alpha,\beta}[\bar{f}_{i}^{\alpha},f_{i}^{\beta}] 
&=& \frac{1}{2}V^{p,0}(w,w')\bar{f}_{iw}^{\alpha}\bar{f}_{iw'}^{\beta}f_{iw'}^{\beta}f_{iw}^{\alpha} \nonumber\\
&+& \frac{1}{4}V^{p,1}(\nu)_{w,w'}\bar{f}_{i,w+\nu}^{\alpha}\bar{f}_{i,-w}^{\alpha}f_{i,-w'}^{\alpha}f_{i,w'+\nu}^{\alpha},\nonumber\\
\end{eqnarray}
 where the impurity full vertex are calculated as discussed in the main text. In general
, the DF vertex $V_{d,i}^{\alpha,\beta}[\bar{f}_{i}^{\alpha},f_{i}^{\beta}]$
contains $n$-body correlation terms introduced by disorder and interaction, but in
the following discussion we will limit ourselves to the leading quartic
term with four external DF fields only.

After taking the derivative with respect to the source field $\eta_{w{\bf k}}$,
the Green function of Eq.~$(\ref{GF_with_S_site})$ reads as 
\begin{equation}
G(w,{\bf k})=\left(\Delta_{w}-\varepsilon_{{\bf k}}\right)^{-1}+\frac{G_d(w,{\bf k})}{\left(\Delta_{w}-\varepsilon_{{\bf k}}\right)^{2}G_{imp}(w)^{2}},\label{origGD}
\end{equation}
 where we define the averaged DF Green function as 
\begin{eqnarray}
G_d(w,{\bf k}) & = & -\lim_{m\rightarrow0}\frac{1}{m}\sum_{\alpha^{\prime}=1}^{m}\int\mathcal{D}\bar{f}\mathcal{D}f\, e^{-\sum_{w{\bf k}\alpha}S_{d0}}\nonumber \\
 & \times & e^{-\sum_{i\alpha\beta w}V_{d,i}^{\alpha,\beta}[\bar{f}_{i}^{\alpha},f_{i}^{\beta}]}f_{w{\bf k}}^{\alpha^{\prime}}\bar{f}_{w{\bf k}}^{\alpha^{\prime}},\label{GD}
\end{eqnarray}
 and $S_{d0}=\bar{f}_{w{\bf k}}^{\alpha}\left[{\displaystyle \frac{(\Delta_{w}-\varepsilon_{{\bf k}})^{-1}+G_{imp}(w)}{G_{imp}^{2}(w)}}\right]f_{w{\bf k}}^{\alpha}$
is the non-interacting DF action.

Notice, that for the case of non-interacting dual fermions when dual
potential is zero, Eq.~$(\ref{origGD})$ reduces to the DMFT+CPA solution
for the lattice Green function with $G(w,{\bf k})=\frac{1}{G_{imp}^{-1}+\Delta_{w}-\varepsilon_{{\bf k}}}.$
Hence, the DMFT+CPA is the zeroth order approximation within our framework.

\section{Dual fermion mapping for the Anderson-Falicov-Kimball model}
Although with different underlying physics, the Anderson-Falicov-Kimball (AFK) model can nevertheless be considered as a simplfied Anderson-Hubbard model by freezing the hopping of one spin flavor of the electrons and serves as a great example to test and verify the dual-fermion approach developed here. In this section, we will give a detailed explanation of the modifications needed to apply the general formalism developed so far onto the AFK model. 

The first apparent difference is the elimination of the hopping term for the f-electrons in the Hamiltonian. Therefore, we can leave 
all the f-electron degree of freedom in the impurity action and only the c-electron degree of freedome needs to be transformed to
the dual-fermion degree of freedom. So the original action can be written as 
\begin{equation}
S=\sum_{i}S_{imp}[\bar{c}_i^{\alpha},c_i^{\alpha}]-\sum_{w{\bf k}\alpha}{\bar{c}_{w{\bf k}}^{\alpha}(\Delta_{w}-\varepsilon_{{\bf k}}-\eta_{w{\bf k}})c_{w{\bf k}}^{\alpha}},
\end{equation}
 with an effective impurity action
\begin{eqnarray}
S_{imp} &=& \sum_{\alpha w}\bar{c}_{wi}^{\alpha}(-iw-\mu+\Delta_{w})c_{wi}^{\alpha}\nonumber \\
 &+& W(\tilde{n}_{i}) + U\sum_{\alpha}\int_{0}^{\beta}d\tau n_{i}^{\alpha}(\tau) n_{i}^{f,\alpha}(\tau).
\end{eqnarray}
Note that different from the Anderson-Hubbard model, there is no hidden spin index dependence in the above action. 
Also note that since the f-electrons only show up in the $S_{imp}$, they will not involve in the dual-fermion mapping.
Due to the similarity of the form of the action $S$ above as compared to Eq. \ref{eq:action_with_imp} for the 
Anderson-Hubbard model, the derivation of dual fermion mapping follows the same. 

The difference in the impurity action affects the parameterization of the dual-fermion system, specifically the dual potential. 
Since the f-electrons are frozen at each site, the scattering processes of c-electrons on the f-electrons are elastic and this 
property will greatly simplify the vertex functions and the dual potential as well. To be more specific, the crossing-symmetric component
of the full vertex function in Eq. \ref{eq:decomp} can be decomposed into two components 
(similar to the decomposition introduced in the reference\cite{janis_afk}) and each of these
two components can be represented as depending on two frequencies only:
\begin{eqnarray}
F^{1}(\nu)_{w,w'} &=& F_{=}(w+\nu,w)\delta_{w,w'}+F_{\parallel}(w,w')\delta_{\nu,0}. 
\end{eqnarray}
Note that the component $F_{=}$ is for the scattering processes which conserve the energy at the horizontal direction,
while the other component $F_{\parallel}$ is for the scattering processes which conserve the energy at the vertical direction.
These two components are related by the crossing symmetry as
\begin{eqnarray}                                                                                                                         
  F_{=}(w,w') = -F_{\parallel}(w,w').                                                     
  \end{eqnarray} 
Here one should remember that this crossing-symmetric component $F^1$ describes all the scattering processes due to the Coulomb interaction
and combined scattering of Coulomb and disorder interactions (type b and type c diagrams in Fig. \ref{fig:F}).

\section{Vertex manipulation}
\label{set:vertex_manipulation} 

The two basic building blocks for constructing the dual fermion diagrams 
are the bare dual Green function and the bare dual vertex. As compared to 
the DF formalism for the clean system, the complexity comes from the vertex
part which requires the differentiation between crossing-asymmetric and crossing-symmetric 
scattering components. These two cannot be treated on the same footing. 
In the following, we will provide a detailed discussion about how to manipulate
the vertex in the calculation of self-energy and vertex functions on the DF lattice.
These equations, such as the Bethe-Salpeter equation and parquet equations, are first
derived from the real fermion system and then generalized to the DF system
by replacing the real fermion quantities by their DF counterparts.

\subsection{Bethe-Salpeter equation}

\begin{figure}[th]
\centerline{ \includegraphics[clip,scale=0.6]{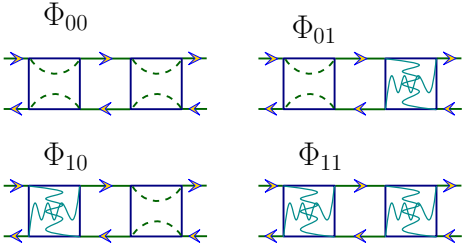}} 
\caption{(Color online). Decomposition of the vertex ladder. Note that $\Phi_{00}$ contributes
to the pure disorder term $F^{0}$ while all the other three contribute
to $F^{1}$.}
\label{fig:phi-bs} 
\end{figure}

The Bethe-Salpeter equation which relates the full vertex $F$ to the irreducible vertex $\Gamma$ reads
\begin{equation}
F(q)_{p,p'} = \Gamma(q)_{p,p'} + \frac{T}{N}\sum_{p''}\Gamma(q)_{p,p''} \chi_{0}(q)_{p''} F(q)_{p'',p'},
\label{BS-eq}
\end{equation}
where each index represents a bundle of Matsubara frequency and momentum indices $p\equiv(iw,\bf{k})$ and $q\equiv (i\nu,\bf{q})$,
and $N$ is the number of sites on the lattice. In the above, we have used the non-perturbative two-particle Green function
for the p-h channel
\begin{equation}
\chi_0(q)_p \equiv G(p+q)G(p).
\end{equation}
Similarly for p-p channel, we have (note the symbol $p$ in the superscript represents p-p channel and should not be confused with
the frequency-momentum index $p$ which appears only in parentheses or subscript)
\begin{equation}
\chi_0^p(q)_p \equiv G(p+q)G(-p).
\end{equation}
To simplify the notation, in the following we will hide the explicit dependence on the indices and write Eq.~\ref{BS-eq} as
\begin{equation}
F = \Gamma + \Gamma \chi_{0} F.
\end{equation}
When solving this equation, the non-trivial part is about how to construct the vertex ladder. 
Since the vertex function can be decomposed into two
components as shown in Eq. \ref{eq:decomp}, we have (remember that ``0" in the superscript represents
the crossing-asymmetric component while ``1" represents crossing-symmetric components, and see Fig. \ref{fig:phi-bs} for the
corresponding Feynman diagrams)
\begin{eqnarray}
\Phi & \equiv & \Gamma\chi_{0}F\nonumber \\
 & = & (\Gamma^{0}+\Gamma^{1})\chi_{0}(F^{0}+F^{1})\nonumber \\
 & = & \Gamma^{0}\chi_{0}F^{0}+(\Gamma^{0}\chi_{0}F^{1}+\Gamma^{1}\chi_{0}F^{0}+\Gamma^{1}\chi_{0}F^{1})\nonumber \\
 & = & \Phi^{0}+\Phi^{1}.
\label{eq:phi} 
\end{eqnarray}
Therefore, the Bethe-Salpeter equation for each component has the
following form
\begin{eqnarray}
F^{0} & = & \Gamma^{0}+\Gamma^{0}\chi_{0}F^{0}\nonumber \\
 & = & [1-\Gamma^{0}\chi_{0}]^{-1}\Gamma^{0}
\end{eqnarray}
and
\begin{eqnarray}
F^{1} & = & \Gamma^{1}+\Gamma^{0}\chi_{0}F^{1}+\Gamma^{1}\chi_{0}F^{0}+\Gamma^{1}\chi_{0}F^{1}\nonumber \\
 & = & [1-\Gamma\chi_{0}]^{-1}\Gamma^{1}[1+\chi_{0}F^{0}].
\end{eqnarray}

\subsection{Parquet equations}

\begin{figure}[th]
\centerline{ \includegraphics[clip,scale=0.55]{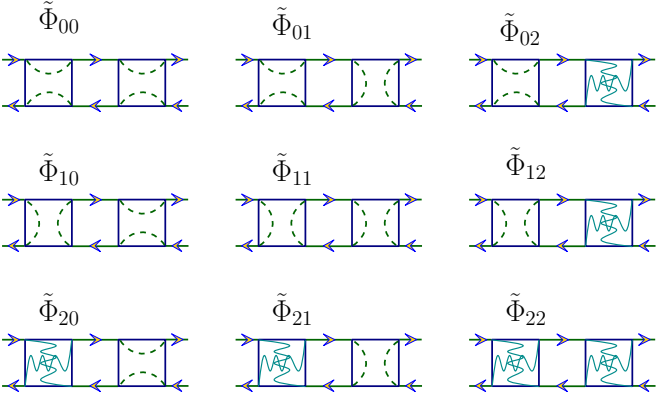}} 
\caption{(Color online). Decomposition of the vertex ladder. Note that $\tilde{\Phi}_{11}$
contains a fermion loop and thus is non-physical. When rotated (see Eq. \ref{eq:rot1}), $\tilde{\Phi}_{00}$
corresponds to the vertical p-h channel and will be canceled
out by the vacuum term. All the left seven terms are physical and
contribute to the p-h irreducible vertex. }
\label{fig:phi} 
\end{figure}

The parquet equations~\cite{s_yang_09} are more involved due to the breaking of the 
crossing symmetry by the crossing-asymmetric component. Since this complexity comes from the
missing of the vertical p-h channel contribution for the crossing-asymmetric component, 
one can therefore pretend there is no such missing diagram and thus the full crossing
symmetries are preserved when constructing the vertex ladders. 
So the parquet equation can be readily written down
and the irreducible vertex can be decomposed into different contributions. 
Only in the very end, the vertical p-h channel contribution for the crossing-asymmetric 
component is removed explicitly to restore the real physical case. By doing
this, one can avoid the possible missing of crossed channel contributions. 

\begin{figure}[th]
\centerline{ \includegraphics[clip,scale=0.7]{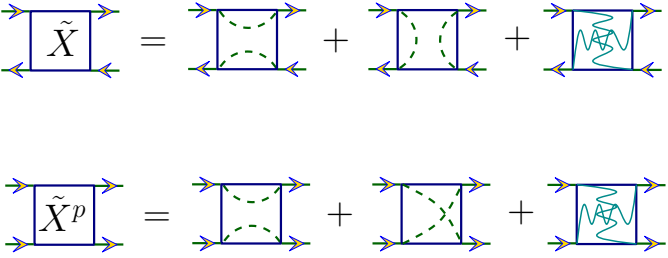}} 
\caption{(Color online). Crossing-symmetrized vertex for both p-h and p-p channels. Note $X$ represents
vertex functions at different levels of reducibility}
\label{fig:vertex} 
\end{figure}

So one has (see Fig. \ref{fig:vertex} for the diagrams)
\begin{equation}
\tilde{F}=F^{0}+F^{0\prime}+F^{1}
\end{equation}
for the full vertex in the p-h channel, where the tilde denotes that
the vacuum term is not subtracted yet and thus contains the p-h vertical
contribution $F^{0\prime}$ from the disorder-only scattering. It would be canceled
out by the vacuum term eventually. Nevertheless, it contributes non-trivially
when constructing the crossed channel contributions. Similarly for
the irreducible p-h vertex

\begin{equation}
\tilde{\Gamma}=\Gamma^{0}+\Gamma^{0\prime}+\Gamma^{1}.
\end{equation}
Then the vertex ladder can be calculated as (see Fig. \ref{fig:phi}) 

\begin{eqnarray}
\tilde{\Phi} & = & (\Gamma^{0}+\Gamma^{0\prime}+\Gamma^{1})\chi_{0}(F^{0}+F^{0\prime}+F^{1}) \nonumber \\
 & = & \Gamma^{0}\chi_{0}F^{0}+\Gamma^{0}\chi_{0}F^{0\prime}+\Gamma^{0}\chi_{0}F^{1} \nonumber\\
 & + & \Gamma^{0\prime}\chi_{0}F^{0}+\Gamma^{0\prime}\chi_{0}F^{0\prime}+\Gamma^{0\prime}\chi_{0}F^{1} \nonumber\\
 & + & \Gamma^{1}\chi_{0}F^{0}+\Gamma^{1}\chi_{0}F^{0\prime}+\Gamma^{1}\chi_{0}F^{1}.
\end{eqnarray}

When rotated (see Eq. \ref{eq:rot1}), the term $\Gamma^{0}\chi_{0}F^{0}$ corresponds
to the pure disorder vertical p-h channel and $\Gamma^{0\prime}\chi_{0}F^{0\prime}$
contains a closed fermion loop, and thus both should be ignored. The
left seven terms are physically meaningful and can be grouped as

\begin{equation}
\Phi_{rot1}=\Phi_{rot1}^{0}+\Phi_{rot1}^{1},
\end{equation}
where

\begin{eqnarray}
\Phi_{rot1}^{0} & = & \left[\Gamma^{0}\chi_{0}F^{0\prime}+\Gamma^{0\prime}\chi_{0}F^{0}
                    +\Gamma^{0\prime}\chi_{0}F^{1}+\Gamma^{1}\chi_{0}F^{0\prime} \right]_{rot1} \nonumber\\
 & = & \left[(\Gamma^{0}+\Gamma^{1})\chi_{0}F^{0\prime}+\Gamma^{0\prime}\chi_{0}(F^{0}+F^{1})\right]_{rot1} \nonumber\\
 & = & \left[\Gamma\chi_{0}F^{0\prime}+\Gamma^{0\prime}\chi_{0}F\right]_{rot1},
\end{eqnarray}
and

\begin{eqnarray}
\Phi_{rot1}^{1} & = & \left[\Gamma^{0}\chi_{0}F^{1}+\Gamma^{1}\chi_{0}F^{0}+\Gamma^{1}\chi_{0}F^{1}\right]_{rot1} \nonumber\\
 & = & \left[(\Gamma^{0}+\Gamma^{1})\chi_{0}F^{1}+\Gamma^{1}\chi_{0}F^{0}\right]_{rot1} \nonumber\\
 & = & \left[\Gamma\chi_{0}F^{1}+\Gamma^{1}\chi_{0}F^{0}\right]_{rot1}.
\end{eqnarray}

The calculation of the crossed p-p channel contribution is straightforward.
Similar to the p-h channel (see Fig. \ref{fig:phi-bs}), the p-p vertex ladder can be calculated as 
\begin{eqnarray}
\Phi^p = \Gamma^p \chi_{0}^p F^p= \Phi^{p,0}+\Phi^{p,1}.
\label{eq:phi-pp} 
\end{eqnarray}
Its rotation is (see Eq. \ref{eq:rot2}) 
\begin{equation}
\Phi_{rot2}^{p}=\Phi_{rot2}^{p,0}+\Phi_{rot2}^{p,1},
\end{equation}
with

\begin{equation}
\Phi_{rot2}^{p,0}=\Gamma^{p,0}\chi_{0}^{p}F^{p,0}\vert_{rot2},
\end{equation}

\begin{equation}
\Phi_{rot2}^{p,1}=\left[\Gamma^{p,0}\chi_{0}^{p}F^{p,1}+\Gamma^{p,1}\chi_{0}^{p}F^{p,0}
                 +\Gamma^{p,1}\chi_{0}^{p}F^{p,1}\right]_{rot2}.
\end{equation}

Therefore, the parquet equations read

\begin{equation}
\Gamma=\Gamma^{0}+\Gamma^{1},
\end{equation}

\begin{equation}
\Gamma^{0}=\Lambda^{0}-\Phi_{rot1}^{0}-\Phi_{rot2}^{p,0},
\end{equation}

\begin{equation}
\Gamma^{1}=\Lambda^{1}-\Phi_{rot1}^{1}-\Phi_{rot2}^{p,1},
\end{equation}
for the p-h channel, and

\begin{equation}
\Gamma=\Gamma^{p,0}+\Gamma^{p,1},
\end{equation}

\begin{equation}
\Gamma^{p,0}=\Lambda^{p,0}+\Phi_{rot3}^{0}-\Phi_{rot2}^{0},
\end{equation}

\begin{equation}
\Gamma^{p,1}=\Lambda^{p,1}+\Phi_{rot3}^{1}-\Phi_{rot2}^{1}
\end{equation}
for the p-p channel. In the above, $\Lambda$ represents the fully irreducible vertex 
for either real fermion system or dual fermion system. For the latter, it might
be approximated by the dual potential $V$. 

The rotations used in the above are defined as~\cite{tam_13, s_yang_09}
\begin{equation}
X(q)_{p,p'}|_{rot1} = X(p'-p)_{p',p+q},
\label{eq:rot1} 
\end{equation}
\begin{equation}
X(q)_{p,p'}|_{rot2} = X(p+p'+q)_{-p',-p},
\label{eq:rot2} 
\end{equation}
\begin{equation}
X(q)_{p,p'}|_{rot3} = X(p-p')_{-p',p+q}.
\label{eq:rot3} 
\end{equation}

\subsection{Schwinger-Dyson equation}

\begin{figure}[th]
\centerline{ \includegraphics[clip,scale=0.6]{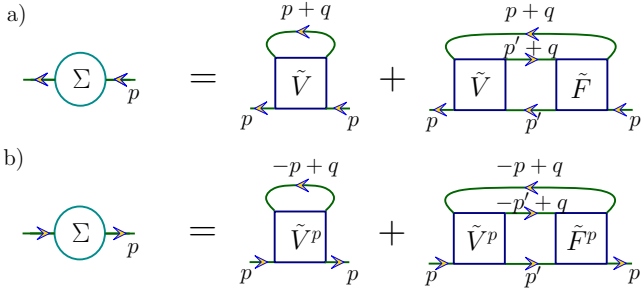}} 
\caption{(Color online). Schwinger-Dyson equation for the dual fermion lattice
expressed through the p-h channel (a) and the p-p channel (b). }
\label{fig:sd} 
\end{figure}

Different from the real fermion lattice, one can use the Schwinger-Dyson equation to construct 
self-energy diagrams efficiently on the dual fermion lattice. For that purpose, one has to use
the crossing-symmetried vertices. Then the Schwinger-Dyson equation from the p-h channel 
(see (a) in Fig. \ref{fig:sd}) reads
\begin{eqnarray}
\Sigma(p) & = & -\frac{T}{N}\sum_{q}\tilde{V}(v)_{w,w}G(p+q) \nonumber\\
          & - &  \frac{T}{2N}\sum_{p',q}\tilde{V}(v)_{w,w'}G(p'+q)G(p')\tilde{F}(q)_{p',p}G(p+q),\nonumber\\
\end{eqnarray}
with the crossing-symmetrized bare p-h vertex defined as (see Fig. \ref{fig:vertex} for the diagrams)
\begin{eqnarray}
\tilde{V}(v)_{w,w} &\equiv&  V^0(w+v,w)\delta_{w,w'} + V^{0\prime}(w,w')\delta_{v,0} \nonumber\\
                   &+&       V^1(v)_{w,w'} \nonumber\\
                   &\equiv&  \tilde{V}^0(v)_{w,w} + V^1(v)_{w,w'}.
\end{eqnarray}
Note that $V=V^0+V^1$ and $V^{0\prime}(w,w')=-V^0(w,w')$.

Or equivalently, it can be written in terms of p-p channel vertices
\begin{eqnarray}
\Sigma(p) & = & \frac{T}{N}\sum_{q}\tilde{V}^p(v)_{w,w}G(-p+q) \nonumber\\
          & - &  \frac{T}{2N}\sum_{p',q}\tilde{V}^p(v)_{w,w'}G(-p'+q)G(p')\tilde{F}^p(q)_{p',p}\nonumber\\
          &   &  \;\;\;\;\;\;\;\;\;\;\times G(-p+q).
\end{eqnarray}
with the crossing-symmetrized bare p-p vertex defined as (see Fig. \ref{fig:vertex} for the diagrams)
\begin{eqnarray}
\tilde{V}^p(v)_{w,w} &\equiv&  V^{p0}(-w+v,w)\delta_{w,w'} + V^{p0\prime}(w',w)\delta_{w+w',v} \nonumber\\
                     &+&        V^{p1}(v)_{w,w'}\nonumber\\
                     &\equiv&  \tilde{V}^{p0}(v)_{w,w} + V^{p1}(v)_{w,w'}.
\end{eqnarray}
Note that $V^p=V^{p0}+V^{p1}$ and $V^{p0\prime}(w',w)=-V^{p0}(w,w')$.

Since the crossing-symmetrized vertices are used, 
a pre-factor $1/2$ is needed for the second term in the above to avoid the double-counting
(the two internal single-particle Green function lines corresponding to indices $p'$ and $p+q$ 
in (a) of Fig. \ref{fig:sd} are indistinguishable and results in this symmetry factor). 
And non-physical diagrams which contain closed fermion loops are produced as well in this way. 
These non-physical diagrams vanish when taking the
replica limit, therefore one has to remove them manually after the construction of self-energy diagrams. 

\subsubsection{First-order contributions}

\begin{figure}[th]
\centerline{ \includegraphics[clip,scale=0.6]{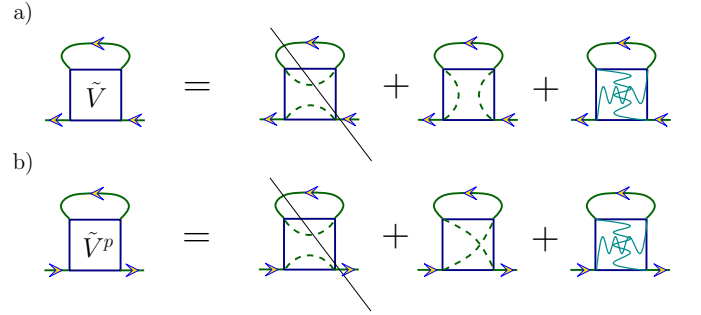}} 
\caption{(Color online). The first-order self-energy diagrams constructed through the Schwinger-Dyson equation
from the p-h channel (a) and p-p channel (b).
Note that the first (Hartree-like) diagram for both channels contains a close fermion loop and is non-physical, 
so one should remove it after the construction.}
\label{fig:sigma1} 
\end{figure}

An example is shown for the first-order diagrams in Fig. \ref{fig:sigma1}. The self-energy can be calculated as
\begin{eqnarray}
\Sigma^{1}_{1st}(w,{\bf k}) & = & -\frac{T}{N}\sum_{v,{\bf k}'}V^{1}(v)_{w,w} G(w,{\bf k}'),
\end{eqnarray}
and
\begin{eqnarray}
\Sigma^{0}_{1st}(w,{\bf k}) & = & \frac{T}{N}\sum_{{\bf k}'}V^{0}(w,w) G(w,{\bf k}'),
\end{eqnarray}
for the crossing-symmetric and crossing-asymmetric vertex components respectively from the p-h channel.
Or equivalently it can be calculated from the p-p channel as
\begin{eqnarray}
\Sigma^{1}_{1st}(w,{\bf k}) & = & \frac{T}{N}\sum_{v,{\bf k}'}V^{p1}(v)_{w,w} G(-w,{\bf k}'),
\end{eqnarray}
and
\begin{eqnarray}
\Sigma^{0}_{1st}(w,{\bf k}) & = & -\frac{T}{N}\sum_{{\bf k}'}V^{p0}(w,w) G(w,{\bf k}').
\end{eqnarray}

Note that after the convergence is achieved,
the first-order contributions should vanish due to the convergence criterion used. 

\subsubsection{Second-order contributions}
\label{sec:DF-2nd} %

\begin{figure}[th]
\centerline{ \includegraphics[clip,scale=0.5]{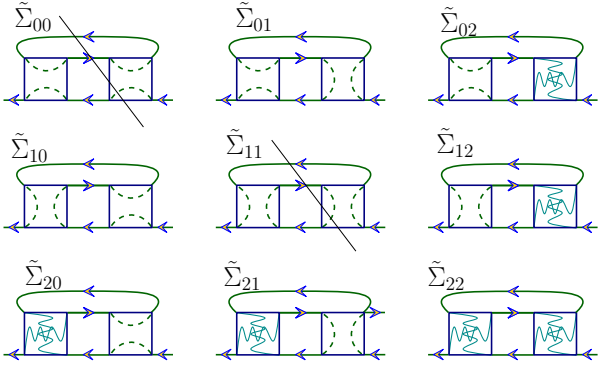}} 
\caption{(Color online). The second-order self-energy diagrams constructed from the Schwinger-Dyson equation
in the p-h channel.
Note that both $\tilde{\Sigma}_{00}$ and $\tilde{\Sigma}_{11}$ contain a close fermion loop and should be removed.}
\label{fig:sigma} 
\end{figure}

For the second-order contributions to the self-energy, one needs to 
approximate the crossing-symmetrized full vertice by the bare one $\tilde{V}$
when using the Schwinger-Dyson equation. One non-trivial 
vertex ladder contribution from the crossing-symmetric component is
\begin{eqnarray}
\Phi^{1}_{2nd} & = & V \bar{\chi}_{0} V - V^{0}\bar{\chi}_{0}V^{0}\nonumber \\
 & = & (V^{0}\bar{\chi}_{0}V^{1}+V^{1}\bar{\chi}_{0}V^{0}+V^{1}\bar{\chi}_{0}V^{1})
\end{eqnarray}
in which the coarse-grained bare two-particle Green function defined as
\begin{eqnarray}
\bar{\chi}_{0}(v,{\bf q})_{w} & = & \frac{1}{N} \sum_{{\bf k}} G(w,{\bf k})G(w+v,{\bf k}+{\bf q}).
\end{eqnarray}
The vertex ladder expression $\Phi = V_1 \bar{\chi}_{0} V_2$ hereafter should be interpreted as the following operation
\begin{eqnarray}
\Phi(v,{\bf q})_{w,w'} = \sum_{w''} V_1(v)_{w,w''} \bar{\chi}_{0}(v,{\bf q})_{w} V_2(v)_{w'',w'}.
\end{eqnarray}
This contribution corresponds to the self-energy diagrams $\tilde{\Sigma}_{02}$, $\tilde{\Sigma}_{20}$ 
and $\tilde{\Sigma}_{22}$ in Fig. \ref{fig:sigma}.
The other contribution which corresponds to self-energy diagrams $\tilde{\Sigma}_{12}$ 
and $\tilde{\Sigma}_{21}$ in Fig. \ref{fig:sigma} is
\begin{eqnarray}
\Phi^{1\prime}_{2nd} & = & V^{0\prime}\bar{\chi}_{0}V^{1}+V^{1}\bar{\chi}_{0}V^{0\prime}.
\end{eqnarray}
The crossing-asymmetric component contributes (corresponding to $\tilde{\Sigma}_{01}$ 
and $\tilde{\Sigma}_{10}$ in Fig. \ref{fig:sigma} )
\begin{eqnarray}
\Phi^{0}_{2nd} & = & V^{0}\bar{\chi}_{0}V^{0\prime} + V^{0\prime}\bar{\chi}_{0}V^{0}.
\end{eqnarray}
The resulting self-energy diagrams are shown in Fig. \ref{fig:sigma}.
Then the self-energy can be calculated as
\begin{eqnarray}
&& \Sigma^{1}_{2nd}(w,{\bf k}) \nonumber\\
& = & -\frac{T}{2} \sum_{v,{\bf q}}
(\Phi^{1}_{2nd}+\Phi^{1\prime}_{2nd})(v,{\bf q})_{w,w}G(w+v,{\bf k}+{\bf q}), \nonumber \\
\end{eqnarray}
and
\begin{eqnarray}
\Sigma^{0}_{2nd}(w,{\bf k}) & = & -\frac{T}{2}\sum_{\bf q}\Phi^{0}_{2nd}(v=0,{\bf q})_{w,w}G(w,{\bf k}+{\bf q}).\nonumber \\
\end{eqnarray}

\begin{figure}[th]
\centerline{ \includegraphics[clip,scale=0.5]{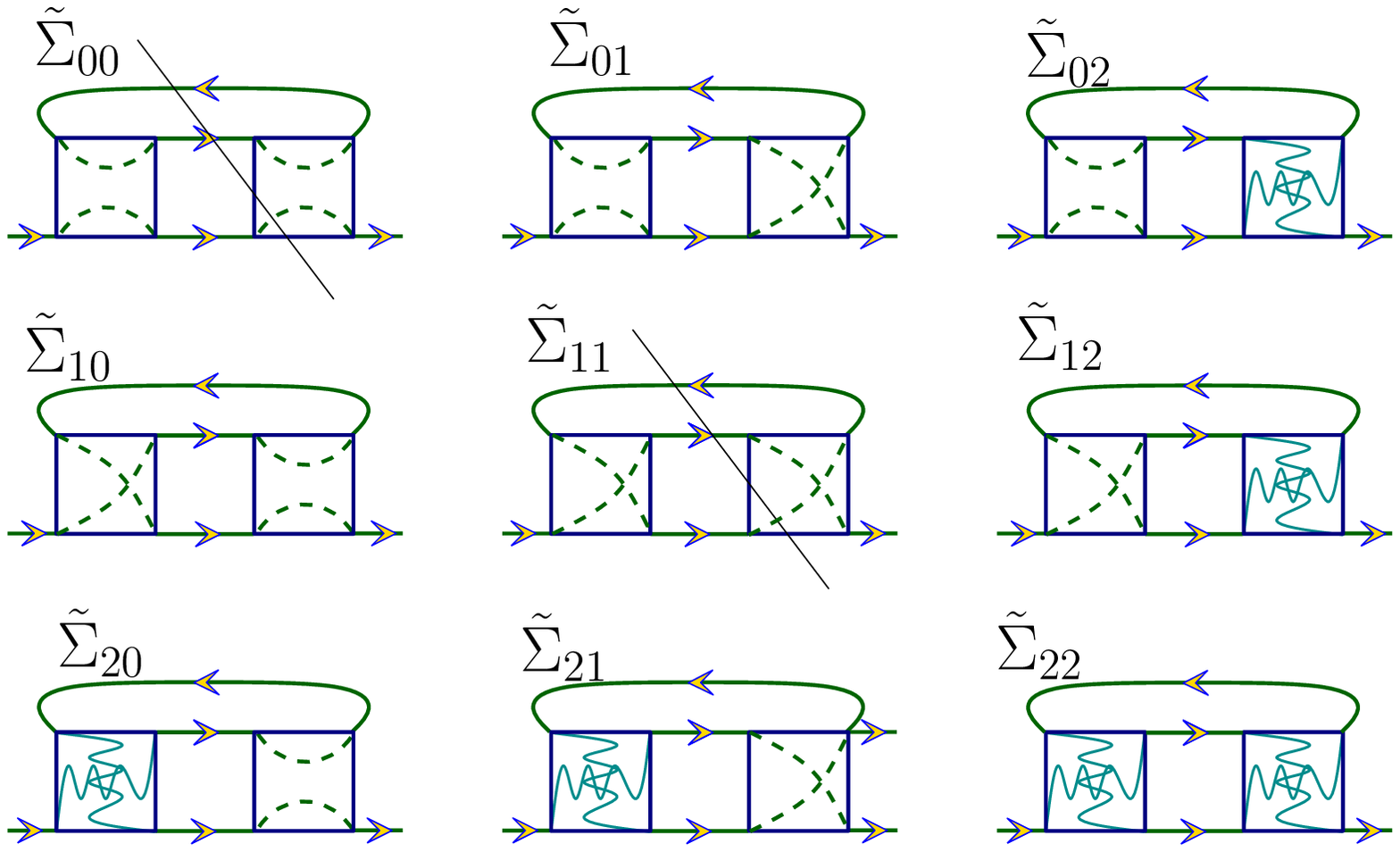}} 
\caption{(Color online). The second-order self-energy diagrams constructed from p-p channel.
Note that both $\tilde{\Sigma}_{00}$ and $\tilde{\Sigma}_{11}$ contain a close fermion loop and should be removed.}
\label{fig:sigma-pp} 
\end{figure}

The second-order self-energy can be equivalently calculated through the p-p channel. 
The vertex ladders needed are
\begin{eqnarray}
\Phi^{p1}_{2nd} & = & {\tilde V}^{p}\bar{\chi}^p_{0}{\tilde V}^{p} - {\tilde V}^{p0}\bar{\chi}^p_{0}{\tilde V}^{p0}  \nonumber \\
 & = & V^{p0}\bar{\chi}^p_{0}V^{p1} + V^{p0\prime}\bar{\chi}^p_{0}V^{p1} + V^{p1}\bar{\chi}^p_{0}V^{p0}\nonumber \\
 & + &V^{p1}\bar{\chi}^p_{0}V^{p0\prime} + V^{p1}\bar{\chi}^p_{0}V^{p1} 
\end{eqnarray}
and 
\begin{eqnarray}
\Phi^{p0}_{2nd} & = & V^{p0}\bar{\chi}^p_{0}V^{p0\prime} + V^{p0\prime}\bar{\chi}^p_{0}V^{p0} \nonumber \\
          & = & 2V^{p0}\bar{\chi}^p_{0}V^{p0\prime}.  
\end{eqnarray}
In the above the coarse-grained bare two-particle Green function in the p-p channel is defined as
\begin{eqnarray}
\bar{\chi}^p_{0}(v,{\bf q})_{w} & = & -\frac{1}{2N} \sum_{{\bf k}} G(w,{\bf k})G(-w+v,-{\bf k}+{\bf q}). \nonumber \\
\end{eqnarray}
in which the symmetry factor $1/2$ is included.
The resulting diagrams are shown in Fig. \ref{fig:sigma-pp} and the self-energy can be calculated as
\begin{eqnarray}
\Sigma^{1}_{2nd}(w,{\bf k}) & = & T \sum_{v,{\bf q}}
\Phi^{p1}_{2nd}(v,{\bf q})_{w,w}G(-w+v,-{\bf k}+{\bf q}), \nonumber \\
\end{eqnarray}
and
\begin{eqnarray}
\Sigma^{0}_{2nd}(w,{\bf k}) & = & T\sum_{\bf q}\Phi^{p0}_{2nd}(v=2w,{\bf q})_{w,w}G(w,-{\bf k}+{\bf q}).\nonumber \\
\end{eqnarray}

To sum up, the second-order self-energy can be calculated as
\begin{eqnarray}
\Sigma_{2nd} & = & \Sigma^{1}_{2nd}  + \Sigma^{1\prime}_{2nd} + \Sigma^{0}_{2nd} \nonumber \\
             & = & \Sigma^{p1}_{2nd} + \Sigma^{p0}_{2nd}.
\end{eqnarray}

\subsubsection{FLEX contributions}
\label{sec:DF-FLEX} %

For the fluctuation-exchange (FLEX) approximation, one should sum
over the ladder diagrams from all the channels.
The p-h channel vertex ladders are
\begin{equation}
\Phi^{1} = [1-V\bar{\chi}_{0}]^{-1}V - [1-V^0\bar{\chi}_{0}]^{-1}V^0,
\end{equation}
\begin{eqnarray}
\Phi^{1\prime} & = & [1-V\chi_{0}]^{-1} V^{0\prime} [1-V\chi_{0}]^{-1} \nonumber \\
               & - & [1-V^0\chi_{0}]^{-1} V^{0\prime} [1-V^0\chi_{0}]^{-1} 
\end{eqnarray}
\begin{eqnarray}
\Phi^{0} & = & [1-V^0\chi_{0}]^{-1} V^{0\prime} [1-V^0\chi_{0}]^{-1} 
\end{eqnarray}
By excluding the second-order contributions, the vertex ladders can be written as
(the index $\alpha$ in the following respresents different components for both p-h and p-p channels)
\begin{eqnarray}
\Phi^{\alpha}_{FLEX} & \equiv & \Phi^{\alpha} - \Phi^{\alpha}_{2nd}
\end{eqnarray}
Then the FLEX self-energy contributions from p-h channel can be calculated as
\begin{eqnarray}
& & \Sigma^{1}_{FLEX}(w,{\bf k}) \nonumber \\
&=& - T \sum_{v,{\bf q}}(\Phi^{1}_{FLEX}+\Phi^{1\prime}_{FLEX})(v,{\bf q})_{w,w}G(w+v,{\bf k}+{\bf q}), \nonumber \\
\end{eqnarray}
and
\begin{eqnarray}
& &\Sigma^{0}_{FLEX}(w,{\bf k}) \nonumber \\
& = & -T\sum_{\bf q}\Phi^{0}_{FLEX}(v=0,{\bf q})_{w,w}G(w,{\bf k}+{\bf q}).\nonumber \\
\end{eqnarray}

The p-p channel vertex ladders are
\begin{equation}
\Phi^{p1} = [1-{\tilde V}^{p}\bar{\chi}_{0}^{p}]^{-1}{\tilde V}^{p} - [1-{\tilde V}^{p0}\bar{\chi}_{0}^{p}]^{-1}{\tilde V}^{p0},
\end{equation}
\begin{equation}
\Phi^{p0} = [1-2V^{p0}\bar{\chi}^p_{0}]^{-1}V^{p0\prime}.
\end{equation}
Then the FLEX self-energy contributions (excluding the second-order contributions) are
\begin{eqnarray}
& &\Sigma^{p1}_{FLEX}(w,{\bf k})  \nonumber \\
& = & T \sum_{v,{\bf q}} \Phi^{p1}_{FLEX}(v,{\bf q})_{w,w}G(-w+v,-{\bf k}+{\bf q}), \nonumber \\
\end{eqnarray}
and
\begin{eqnarray}
& &\Sigma^{p0}_{FLEX}(w,{\bf k})  \nonumber \\
& = & T\sum_{\bf q}\Phi^{p0}_{FLEX}(v=2w,{\bf q})_{w,w}G(w,-{\bf k}+{\bf q}).\nonumber \\
\end{eqnarray}

The FLEX self-energy is calculated by summing all these contributions 
\begin{eqnarray}
\Sigma_{FLEX} & = & \Sigma^{1}_{FLEX}+\Sigma^{1\prime}_{FLEX}+\Sigma^{0}_{FLEX} \nonumber \\
              & + & \Sigma^{p1}_{FLEX} + \Sigma^{p0}_{FLEX} +  \Sigma_{2nd}.
\end{eqnarray}

\section{Calculation of dc conductivity}

We calculate the dc conductivity as~\cite{Denteneer, h_terletska_13}
\begin{equation}
\sigma_{dc}=\frac{\beta^2}{\pi}\chi_{xx}({\bf q}=0,\tau=\frac{\beta}{2}),
\end{equation}
with the current-current correlation function $\chi_{xx}=\langle j_x({\bf q},\tau) j_x(-{\bf q},0) \rangle$,
and $\beta=1/T$ the inverse temperature.
The current-current correlation function can be Fourier transformed from the frequency space
\begin{equation}
\chi^{xx}({\bf q}=0,\tau=\frac{\beta}{2})=T\sum_{i\nu_{m}}e^{-i\nu_{m}\frac{\beta}{2}}\chi^{xx}({\bf q}=0,i\nu_{m}),
\end{equation}
and then it can be related with the two-particle Green function
\begin{eqnarray}
 &  & \chi^{xx}({\bf q}=0,\nu\equiv i\nu_{m})\nonumber \\
 & = & \frac{T}{N^{2}}\sum_{w,w^{\prime};{\bf k},{\bf k}^{\prime}}\chi^{xx}({\bf q}=0, \nu)_{w+\nu,{\bf k};w^{\prime},{\bf k}^{\prime}}\nonumber \\
 & = & -\frac{T}{N}\sum_{w;{\bf k}}v_{{\bf k}}^{2}\chi_{0}({\bf q}=0,\nu)_{w,{\bf k}}\nonumber \\
 & - & \frac{T^{2}}{N^{2}}\sum_{w,w^{\prime};{\bf k},{\bf k}^{\prime}}v_{{\bf k}}\chi_{0}({\bf q}=0,\nu)_{w,{\bf k}}F^{rf}({\bf q}=0,\nu)_{w,{\bf k};w^{\prime}, {\bf k}^{\prime}}\nonumber \\
 &&    \times \chi_{0}({\bf q}=0,\nu)_{w',{\bf k}'} \,v_{{\bf k}^{\prime}}.
\end{eqnarray}
The full vertex $F^{rf}$ is defined on the real fermion space, and has to be mapped from its dual fermion counterpart through
\begin{equation}
F^{rf}(q)_{p,p^{\prime}}=S(p+q)S(p)F(q)_{p,p^{\prime}}S(p^{\prime}+q)S(p)
\end{equation}
with the assistance of the transformation matrix defined as
\begin{eqnarray}
S & = &  -\frac{1}{1+G_{imp}\Sigma_{d}}.
\end{eqnarray}
Now one can employ the Bethe-Salpeter and parquet equations discussed in Appendix \ref{set:vertex_manipulation} to take into account all the crossed channel contributions 
for the dual fermion full vertex $F$.
To accelerate the convergence of the calculation of the conductivity on the dual fermion lattice, the embedding scheme proposed recently~\cite{df-embedding}
is employed through the calculations.

\bibliography{df_afk}

\end{document}